\documentclass[sigconf]{acmart}
\usepackage{amsmath,amsfonts}
\usepackage{graphicx}
\usepackage{pgf}
\usepackage{xcolor}
\usepackage{mathtools}
\usepackage{xspace}
\usepackage{array}
\usepackage[utf8]{inputenc}
\usepackage{longtable}
\usepackage{algorithm}  
\usepackage{algpseudocode}  
\usepackage{tikz}
\usetikzlibrary{fit,arrows,shapes,snakes,automata,backgrounds,petri}
\usetikzlibrary{arrows.meta}
\usepackage{multirow}
\usepackage{listings}
\usepackage{caption}
\usepackage{fancyvrb}
\usepackage{bbding}
\usepackage{pifont}
\usepackage{wasysym}
\usepackage{tabularx}
\usepackage{pdfpages}
\usepackage{colortbl}
\usepackage{enumitem}
\usepackage{pifont}
\usepackage{makecell}
\usepackage{comment}
\usepackage{booktabs}
\usepackage{textcomp}
\usepackage{lscape}
\usepackage{verbatim}
\usepackage{textcomp}
\usepackage[most]{tcolorbox}
\usepackage{subcaption}
\usepackage{hyperref}
\usepackage{soul}

\usepackage[normalem]{ulem}

\AtBeginDocument{%
  \providecommand\BibTeX{{%
    \normalfont B\kern-0.5em{\scshape i\kern-0.25em b}\kern-0.8em\TeX}}}

\copyrightyear{2024}
\acmYear{2024}
\setcopyright{acmlicensed}\acmConference[CCS '24]{Proceedings of the 2024 ACM SIGSAC Conference on Computer and Communications Security}{October 14--18, 2024}{Salt Lake City, UT, USA}
\acmBooktitle{Proceedings of the 2024 ACM SIGSAC Conference on Computer and Communications Security (CCS '24), October 14--18, 2024, Salt Lake City, UT, USA}
\acmDOI{10.1145/3658644.3690312}
\acmISBN{979-8-4007-0636-3/24/10}

\newcommand*{\mybox}[1]{\framebox{#1}}

\algrenewcommand\algorithmicindent{0.5em}

\newcommand{\cmark}{\ding{51}}%
\newcommand{\xmark}{\ding{55}}%

\newcommand{\system}{\text{\fontfamily{lmtt}\selectfont Proteus}\xspace}
\newcommand{\ltlparser}{\ensuremath{\mathsf{RegExGenerator}}\xspace}
\newcommand{\dpalgo}{\ensuremath{\mathsf{TraceBuilder}}\xspace}
\newcommand{\dispatcher}{\ensuremath{\mathsf{TraceDispatcher}}\xspace}

\newcommand{\pmodel}{\ensuremath{\mathcal{M}}\xspace}

\newcommand{\pimp}{\ensuremath{\mathcal{I}_P}\xspace}

\newcommand{\ie}{\emph{i.e.}\xspace}
\newcommand{\eg}{\emph{e.g.}\xspace}

\newcommand{\NumLTEDevice}{11 }

\newcommand{\NumLTEVulnerabilities}{10 }

\newcommand{\NumLTENew}{3 }
\newcommand{\NumLTEVulInstanece}{31 }

\newcommand{\NumBLEDevice}{12 }

\newcommand{\NumBLEVulnerabilities}{15 }

\newcommand{\NumBLENew}{7 }
\newcommand{\NumBLEVulInstanece}{81 }

\newcommand{\protomachine}{\ensuremath{\mathcal{M}}\xspace}
\newcommand{\protoproperty}{\ensuremath{\varphi}\xspace}
\newcommand{\budget}{\ensuremath{\beta}\xspace}
\newcommand{\transitionRel}{\ensuremath{\mathcal{R}}\xspace}
\newcommand{\trace}{\ensuremath{\pi}\xspace}
\newcommand{\tracesub}[1]{\ensuremath{\pi_{#1}}\xspace}
\newcommand{\tracesup}[1]{\ensuremath{\pi^{#1}}\xspace}

\newcommand{\TotalVul}{\pgfmathparse{int(round(\NumLTEVulnerabilities+\NumBLEVulnerabilities))}\pgfmathresult\xspace}

\newcommand{\VulAcked}{14\xspace}

\newcommand{\enablesone}{\footnotesize\textsc{enable\_s1}\normalsize\xspace}
\newcommand{\attachreq}{\footnotesize\textsc{attach\_request}\normalsize\xspace}
\newcommand{\idreq}{\footnotesize\textsc{identity\_request}\normalsize\xspace}

\newcommand{\idreqtmsi}{\footnotesize\textsc{identity\_ request (TMSI)}\normalsize\xspace}
\newcommand{\idrespimsi}{\footnotesize\textsc{identity\_ response (IMSI)}\normalsize\xspace}

\newcommand{\authreq}{\footnotesize\textsc{authentication\_request}\normalsize\xspace}

\newcommand{\authresp}{\footnotesize\textsc{authentication\_response}\normalsize\xspace}
\newcommand{\authfail}{\footnotesize\textsc{authentication\_failure}\normalsize\xspace}
\newcommand{\emmstatus}{\footnotesize\textsc{emm\_status}\normalsize\xspace}
\newcommand{\smcmd}{\footnotesize\textsc{security\_mode\_command}\normalsize\xspace}
\newcommand{\smcmdreplay}{\footnotesize\textsc{security\_mode\_command:replay == 1}\normalsize\xspace}

\newcommand{\smrej}{\footnotesize\textsc{security\_ mode\_reject}\normalsize\xspace}
\newcommand{\gutiral}{\footnotesize\textsc{guti\_reallocation\_command}\normalsize\xspace}
\newcommand{\gutiralrep}{\footnotesize\textsc{guti\_reallocation\_command: replay == 1}\normalsize\xspace}
\newcommand{\gutiralc}{\footnotesize\textsc{guti\_reallocation\_complete}\normalsize\xspace}
\newcommand{\rrcsmcmd}{\footnotesize\textsc{rrc\_ security\_ mode\_command}\normalsize\xspace}
\newcommand{\rrcsmcmplt}{\footnotesize\textsc{rrc\_security\_ mode\_ complete}\normalsize\xspace}
\newcommand{\countercheck}{\footnotesize\textsc{counter\_check}\normalsize\xspace}

\newcommand{\attachacc}{\footnotesize\textsc{attach\_accept}\normalsize\xspace}
\newcommand{\attachcomp}{\footnotesize\textsc{attach\_complete}\normalsize\xspace}

\newcommand{\nullaction}{\footnotesize\textsc{null\_action}\normalsize\xspace}

\newcommand{\smcomplete}{\footnotesize\textsc{security\_mode\_complete}\normalsize\xspace}

\newcommand{\smcmdintegrityzero}{\footnotesize\textsc{security\_mode\_command: integrity == 1 \& EIA == 1 \& security\_header == 3}\normalsize\xspace}
\newcommand{\zerointegrity}{\footnotesize\textsc{integrity == 0}\normalsize\xspace}
\newcommand{\eiaone}{\footnotesize\textsc{EIA == 1}\normalsize\xspace}
\newcommand{\shthree}{\footnotesize\textsc{security\_header == 3}\normalsize\xspace}

\newcommand{\scanreq}{\footnotesize\textsc{scan\_req}\normalsize\xspace}
\newcommand{\scanresp}{\footnotesize\textsc{scan\_resp}\normalsize\xspace}
\newcommand{\connectionreq}{\footnotesize\textsc{connection\_ request}\normalsize\xspace}
\newcommand{\versionreq}{\footnotesize\textsc{version\_ request}\normalsize\xspace}
\newcommand{\versionreqextllzero}{\footnotesize\textsc{version\_request:ext\_ll == 0}\normalsize\xspace}
\newcommand{\versionresp}{\footnotesize\textsc{version\_request:ext\_ll == 0}\normalsize\xspace}

\newcommand{\mtureq}
{\footnotesize\textsc{MTU\_request}\normalsize\xspace}
\newcommand{\encpausereq}{\footnotesize\textsc{encryption\_pause\_request}\normalsize\xspace}

\newcommand{\encpauseresp}{\footnotesize\textsc{encryption\_pause\_response}\normalsize\xspace}

\newcommand{\lengthreq}{\footnotesize\textsc{length\_request}\normalsize\xspace}
\newcommand{\pairreq}{\footnotesize\textsc{pairing\_request}\normalsize\xspace}
\newcommand{\pairresp}{\footnotesize\textsc{pairing\_response}\normalsize\xspace}
\newcommand{\keyexchange}{\footnotesize\textsc{public\_key\_exchange}\normalsize\xspace}

\newcommand{\pauseencreqplaintext}{\footnotesize\textsc{enc\_pause\_req\_plaintext}\normalsize\xspace}

\newcommand{\pairconfirm}{\footnotesize\textsc{sm\_confirm}\normalsize\xspace}

\newcommand{\pairrandom}{\footnotesize\textsc{sm\_random}\normalsize\xspace}
\newcommand{\encreq}{\footnotesize\textsc{encryption\_request}\normalsize\xspace}
\newcommand{\encresp}{\footnotesize\textsc{encryption\_response}\normalsize\xspace}
\newcommand{\pauseencreq}{\footnotesize\textsc{pause\_ encryption\_request}\normalsize\xspace}
\newcommand{\pauseencresp}{\footnotesize\textsc{pause\_ encryption\_response}\normalsize\xspace}
\newcommand{\dhkeycheck}{\footnotesize\textsc{dh\_key\_check}\normalsize\xspace}

\newcommand{\chmreq}{\footnotesize\textsc{channel\_map\_request}\normalsize\xspace}
\newcommand{\prelimexampleone}{\footnotesize\textsc{[scan\_req/ scan\_resp, con\_req/ null\_action, version\_req: ext\_ll==0/ version\_resp, pair\_req: sc==1/ pair\_resp, key\_exchange / key\_ response]}\normalsize\xspace}

\newcommand*{\priority}[1]{%
   \begin{tikzpicture}[scale=0.12]
      \draw (0,0) circle (1);
      \fill[opacity=1,fill=black] (0,0) -- (90:1) arc (90:90-#1*3.6:1) -- cycle;
   \end{tikzpicture}
}

\newcommand{\categorythree}{\priority{100}}
\newcommand{\categorytwo}{\priority{50}}
\newcommand{\categoryone}{\priority{0}}

\begin{document}

\title{State Machine Mutation-based Testing Framework for Wireless Communication Protocols}

%\author{Anonymous Authors}
% \affiliation{%
\author{Syed Md Mukit Rashid}
\affiliation{%
  \institution{The Pennsylvania State University}
  \city{University Park}
  \state{PA}
  \country{United States}}
\email{szr5848@psu.edu}

\author{Tianwei Wu}
\affiliation{%
  \institution{The Pennsylvania State University}
  \city{University Park}
  \state{PA}
  \country{United States}}
\email{tvw5452@psu.edu}

\author{Kai Tu}
\affiliation{%
  \institution{The Pennsylvania State University}
  \city{University Park}
  \state{PA}
  \country{United States}}
\email{kjt5562@psu.edu}

\author{Abdullah Al Ishtiaq}
\affiliation{%
  \institution{The Pennsylvania State University}
  \city{University Park}
  \state{PA}
  \country{United States}}
\email{abdullah.ishtiaq@psu.edu}

\author{Ridwanul Hasan Tanvir}
\affiliation{%
  \institution{The Pennsylvania State University}
  \city{University Park}
  \state{PA}
  \country{United States}}
\email{rpt5409@psu.edu}

\author{Yilu Dong}
\affiliation{%
  \institution{The Pennsylvania State University}
  \city{University Park}
  \state{PA}
  \country{United States}}
\email{yiludong@psu.edu}

\author{Omar Chowdhury}
\affiliation{%
  \institution{Stony Brook University}
  \city{Stony Brook}
  \state{NY}
  \country{United States}}
\email{omar@cs.stonybrook.edu}

\author{Syed Rafiul Hussain}
\affiliation{%
  \institution{The Pennsylvania State University}
  \city{University Park}
  \state{PA}
  \country{United States}}
\email{hussain1@psu.edu}
% \affiliation{%

% \affiliation{%

% \affiliation{%

% \affiliation{%

% \affiliation{%

% \affiliation{%

% \affiliation{%

%\renewcommand{\shortauthors}{Syed et al.}
\renewcommand{\shortauthors}{Syed Md Mukit Rashid et al.}
%% No italics, no superscripts
%% Use footnote or author note to identify equal contribution and/or contact author info

\begin{abstract}
  This paper proposes \system, a 
protocol state machine, property-guided, and budget-aware 
automated testing approach for discovering  
logical vulnerabilities in wireless protocol implementations. 
\system maintains its budget awareness by generating test 
cases (\ie, each being a sequence of protocol messages) that are not only 
\emph{meaningful} (\ie, the test case mostly follows the 
desirable protocol flow except for some controlled 
deviations) but also have a high probability of violating  
the desirable properties. 
{
To demonstrate its effectiveness, we evaluated \system in two different protocol implementations, namely 4G LTE and BLE, across 23 consumer devices (\NumLTEDevice for 4G LTE and \NumBLEDevice for BLE). %During its 
%evaluation, 
\system discovered \TotalVul unique issues, including 112 instances. 
%Of these \TotalVul, \NewVul are newly discovered, and 16 are previously known but found in new devices not previously reported to have the vulnerability. 
Affected vendors have positively acknowledged \VulAcked vulnerabilities through 5 CVEs. %or bug bounty. 
%\syed{10 vs. 12}
}

% (4 + 1 + 4)9 + (8 + 2 + 12 + 2 + 4 + 6 + 10) = 44 - vuln. instances
% known but found in new devices - 

\end{abstract}






%\settopmatter{printfolios=true} 
%TODO: need to comment out before submission

\maketitle

\section{Introduction}
\label{sec:introduction}

Due to the pervasiveness and broadcast-based over-the-air (OTA) communication, 
wireless protocols (\eg, BLE, LTE, Wi-Fi) are attractive 
targets for attackers, especially because vulnerabilities in them can be 
used as a stepping stone by adversaries to launch attacks against applications relying on them. 
Software testing has proven to be the most effective and dominant approach for ensuring the correctness 
of wireless protocol implementation by uncovering bugs in the pre-deployment stages. 
% Software testing is the predominant approach for uncovering bugs in wireless protocol implementations. 
This paper focuses on designing, 
developing and evaluating  an automated 
%, model- and property-guided, budget-aware automated 
testing approach called \system for uncovering  \emph{logical/semantic vulnerabilities}
in wireless protocol implementations~\cite{lteinspector, hussain2019insecure, rupprecht2018security, 5greasoner, profactory, blesa, wu2022formal}. 
The class of logical bugs we focus on is only those that induce the 
implementation to deviate from 
the intended design \emph{without} necessarily causing a crash. 
Such logical bugs are not only challenging to discover but also often have severe security and 
privacy implications (e.g., authentication bypass) and go undiscovered during pre-deployment stage 
testing. These classes of bugs  
are the focus of this paper.

Any testing approaches focusing on discovering such logical bugs in commercial-off-the-shelf (COTS) 
wireless protocol implementations must be aware of the following salient aspects. 
\ding{172} COTS implementations %in these devices  
are typically closed-source and expose only their input-output interfaces, making them only amenable to black-box testing. 
\ding{173} Wireless protocols are often stateful. Consequently, triggering 
a vulnerability %in a target implementation 
amounts to injecting the right protocol packet type and payload \emph{only when} the protocol is in a specific internal protocol state. 
In addition, driving the implementation to the bug-triggering state may require a long sequence of protocol packets. 
%, which, in turn, may require a long sequence of packets to drive the protocol to the bug-triggering state; 
\ding{174} When executing a test case, the target device's protocol under test has to be in the initial 
state. After running each test case, the device thus has to be reset. 
This results in a very high and fixed amortized cost of running a single test case over-the-air 
(e.g., $\sim\!\!1.5$ minute for BLE),  limiting the number of test cases executable within a given time \textit{budget}. 
%Since the cost of running an OTA query and the total number of executable test cases are confined and 
%In addition, since test coverage is unavailable in black-box settings, 
% Hence, one needs to carefully adjust mutations and lengths for the test cases according to the available time. 
Given a fixed testing budget, it is therefore crucial that the test cases used are not wasted and indeed meaningful, that is, they 
have a high probability of exercising the bug-inducing behavior in the implementations.
%This results in the OTA execution of a test case taking a significant time (e.g., $\sim\!\!1.5$ minute for BLE), limiting the number of test cases executable within a given \textit{budget}.

% Although some prior work developed automated methods to test network protocol 
% implementations (e.g., TCP, UDP, and TLS), they are inadequate for wireless protocol 
% implementations in COTS devices due to their dependency on source code availability ~\cite{}. 
Existing research efforts that specialize in black-box testing of wireless communication protocol implementations  %to detect logical flaws 
can be categorized into the following high-level categories: 
\textbf{(A)} \textit{Manual analysis or fixed test case-based approaches}~\cite{park2016white, antonioli2022blurtooth,Shaik2016,SeriVishnepolsky2017}; 
\textbf{(B)} \textit{Reverse engineering-based approaches} ~\cite{hernandez2022firmwire, basespec, basesafe, baseband-1, baseband-2, ruge2020frankenstein}; 
\textbf{(C)} \textit{State machine learning-based approaches}~\cite{dikeue, blediff, chlosta2021challenges, fingerprinting-ble, sweyntooth}. 
Approaches in categories (A) and (B) are either unscalable due to 
 manual effort or ineffective in 
identifying intricate bugs in complex and stateful protocols that require long execution packet traces to be exercised. 
Category (C) approaches can address 
both aspects \ding{172} and \ding{173}, but fail to address \ding{174}. 
Among the category (C) approaches the ones that rely on 
automata learning~\cite{dikeue, blediff} need to run a large 
number of over-the-air (OTA) queries to obtain the protocol state machine before 
testing can commence, hence their testing-budget agnosticism. 

% Beyond these three high-level categories of approaches, 
% other approaches either require manual efforts to guide the testing approach ~\cite{dtls_fuzz, sweyntooth, dy_fuzzing} 
% or require access to source code ~\cite{afl}.
% there are 
% several other approaches ~\cite{dtls_fuzz, sweyntooth, dy_fuzzing}, 
% which  
% or require access to source code. 
Furthermore, many of the above approaches rely on 
\emph{differential testing}, in which diverse implementations-under-test (IUT)
are used as cross-checking \emph{test oracles} to find logical vulnerabilities. 
However, these oracles are inherently unfaithful 
as test oracles because they all can suffer from the same logical vulnerability. 
% {Finally, most fuzzing techniques ~\cite{feng2021snipuzz,lyu2019mopt,gan2018collafl} attempt to optimize their approach to maximize vulnerability discovery, considering an \textit{implicit} testing budget. However, since they do not consider the testing budget explicitly, they cannot fully adapt their fuzzing approach according to the available testing budget.} \emph{In summary, none of the approaches adapt their behavior to the given testing budget, making them inadequate in our context}. 
%
{Finally, most fuzzing techniques ~\cite{feng2021snipuzz,lyu2019mopt,gan2018collafl} only 
\emph{implicitly} consider a testing budget. They aim to effectively use the given testing budget in terms 
of the number of vulnerabilities discovered by attempting to minimize the execution time of 
each test case while also taking guidance from some rich forms of coverage 
information (e.g., code coverage). However, in a testing setup like ours, where the cost of 
running each test case cannot be substantially minimized and the availability of coverage 
information is limited, the existing philosophy of decreasing the execution time of each test case 
cannot be effectively adopted.  
% consider only an \textit{implicit} testing budget and aim to minimize the execution time to maximize the total number of test cases and code coverage to maximize vulnerability discovery. 
% Since they do not consider the testing budget explicitly, they cannot fully adapt their fuzzing approach, e.g., the number of mutations according to the available testing budget.}
%\emph{In summary, none of the existing approaches are directly applicable to our existing testing setup, and a new testing philosophy is warranted}. 
\emph{In summary, none of the current approaches suit our testing setup and thus warrant a new testing philosophy.}

We address the above limitations by designing  
an adaptive testing-budget-aware, stateful, black box testing approach 
called \system for COTS wireless protocol implementations. 
Concretely, \system{}'s testing philosophy takes advantage of the 
following three observations. 
\ding{182} Since COTS devices generally pass through a quality assurance stage where they are tested 
for conformance and interoperability, undiscovered bugs are more likely to occur when implementations subtly deviate from rare but good protocol flows. As such, a test case is \emph{meaningful} if it \emph{mostly} adheres to a desired protocol message sequence (\eg, sending protocol messages 
only after the initial connection request message), except for some controlled perturbations. \ding{183} Any testing-budget-aware 
approach will be effective only if it reduces test case wastage by generating \emph{only}  
meaningful test cases that have a high probability of triggering logical vulnerabilities within the given budget. \ding{184} The number of possible meaningful test cases depends on the amount of perturbation and the maximum length we allow. We can adapt an approach to maximize vulnerability discovery within a given testing budget if they can be explicitly controlled.

% \ding{182} Any testing-budget-aware 
% approach will be effective only if it reduces test case wastage by generating \emph{only}  
% meaningful test cases that have a high probability of triggering  
% logical vulnerabilities within the given budget. 
%  In contrast, a test case is 
% considered  \emph{meaningful} if it \emph{mostly} adheres to a desired protocol 
% message sequence (\eg, sending protocol messages 
% only after the initial connection request message), except for some controlled perturbations. 
% \ding{183} Since COTS devices generally pass through a quality assurance stage where they are tested 
% for conformance and interoperability, undiscovered bugs are more likely to occur when 
% implementations subtly deviate from rare, but good protocol flows. 

 \system takes three inputs for generating test cases, namely, 
a guiding protocol state machine \protomachine, 
a set of desirable security and privacy properties $\Phi$ expressed in past-time 
linear temporal logic formulae, and a testing budget \budget{}. \system relies on \protomachine, either derived from the standard ~\cite{rfcnlp, hermes} 
or extracted from an implementation ~\cite{chiron, statelifter}, to ensure that the generated test cases are indeed meaningful. 
%\protomachine can be either derived from the standard ~\cite{rfcnlp, hermes} 
%or extracted from an implementation ~\cite{chiron, statelifter}. 
\system{} generates test cases through controlled perturbation over \protomachine, thus \emph{mostly} 
capturing good protocol behavior (realization of observation \ding{182}). 
Similarly, any property \protoproperty of the protocol under test guides \system{}
to only generate test cases likely to trigger violation of \protoproperty (realization of observation \ding{183}). 
The set of properties $\Phi$ serves not only as a faithful 
test oracle but also liberates \system from needing to have access to diverse 
implementations of the same protocol to test a single protocol implementation.  
Furthermore, the number of test cases to be generated by \system is \emph{explicitly} controlled by the 
testing budget parameter \budget. It achieves budget-awareness
by controlling the maximum size of a test case (\ie, the length of the message sequence) 
and the number of perturbations to be applied in a good protocol flow for generating test cases (realization of observation \ding{184}). In addition, as discussed in \ding{182}, the number of mutations one needs to consider (given by \budget) 
need not be large. This observation is corroborated by our findings, where most vulnerabilities 
are discovered by considering only \textbf{2} mutations of \protomachine.

Conceptually, \system{}' design is inspired by mutation-based testing, where the 
guiding protocol state machine \protomachine is 
%intentionally 
mutated to obtain another state machine $\protomachine^\ast$ such that $\protomachine^\ast$ violates \protoproperty. If the implementation under test is equivalent to $\protomachine^\ast$, 
we can automatically obtain the vulnerability's root cause by tracking the mutation applied to $\protomachine$ for obtaining $\protomachine^\ast$. 
\system realizes this conceptual design through a novel three-stage 
approach. First, \system uses a novel algorithm to automatically synthesize test case templates that are guaranteed to violate \protoproperty. 
Second, \system{} instantiates these test case templates using a dynamic programming algorithm for the given \protomachine, \protoproperty and 
\budget (\ie, the maximum length of the message sequence and the number of allowed mutations from \protomachine). 
Finally, \system{} efficiently schedules, concretizes, and executes the instantiated test cases OTA. \system{} analyzes the output generated 
by the implementation to discover logical vulnerabilities.

To demonstrate the efficacy of \system, we have tested several protocol implementations of two popular wireless communication protocols: 4G LTE cellular network and Bluetooth Low Energy (BLE). For 4G LTE, we have tested \NumLTEDevice devices and identified \NumLTEVulnerabilities{} issues, including \NumLTENew new ones. For BLE, we have tested \NumBLEDevice devices and identified \NumBLEVulnerabilities issues, including \NumBLENew new ones. 

% \textbf{Contributions.} 
In summary, this paper makes the following contributions. %of our paper is as follows. 
\begin{itemize}[noitemsep,topsep=0pt,leftmargin=0.4cm]
    \item We developed \system, which is an efficient, 
a black box, protocol state machine and property-guided, budget-aware 
automated testing approach for discovering 
logical vulnerabilities in wireless protocol implementations.    
    \item We developed two novel algorithms that cooperatively 
generate meaningful test case templates that are likely to violate 
a given set of security properties under the guidance of a state machine. 
% closely adhere 
% to a given protocol state machine while violating the desired properties. Although 
% developed in the context of \system, the algorithm is of general interest for 
% stateful software testing. 
    \item We developed an effective dispatcher that efficiently schedules 
test cases for maximizing the number of property violations within the allocated testing budget, issues OTA test cases to the devices and analyzes the output to find logical vulnerabilities. 
% It leverages \system's capability to generate property-guided queries, OTA response of the target implementation, and queries executed thus far.
    % \item We developed \ltlparser, which efficiently analyzes a given security property and builds REs representing a pattern of message sequences interesting to test.
    \item We evaluated \system on LTE and BLE to determine its efficacy. It identified \NumLTENew new issues in \NumLTEDevice LTE implementations and \NumBLENew new issues testing on \NumBLEDevice BLE implementations.
\end{itemize}

\noindent\textbf{Responsible disclosure.}
% \textcolor{blue}{
We have responsibly reported all of our new findings to the affected vendors. For BLE, the vendors acknowledged 11 issues and assigned 3 CVEs; for LTE, the vendors acknowledged 3 issues with 2 CVEs. 
%We also obtained a \$3,000 bug bounty for one of the LTE vulnerabilities and \$2,850 bug bounty for two BLE vulnerabilities. 
{
The source code of \system is available at ~\cite{proteus_code}.
%A full extended version of our paper, including Appendix, can be found in ~\cite{proteus_extended}.
%after the expiration of the vendor-requested embargo.
}

%\syed{Update responsible disclosure.}

% Now, security properties can generally be expressed in the form of an LTL formula, and violations of these properties are generally sequences that follow a form of regular expression(s). To automatically generate regular expressions of sequences that violate a particular property, \system also proposes another component \ltlparser, which builds a syntax tree from the LTL formula and performs a heuristic algorithm to generate possible regular expressions of sequences that violate the given LTL property. \dpalgo later leverages 
%\vspace{-0.2cm}   
\section{Preliminaries and Notations}
\label{sec:preliminaries}
% In this section, we review the preliminary concepts and notations necessary to understand 
% our technical contributions. 

\noindent\textbf{Protocol state machine (PSM).}
A protocol model or state machine (PSM) \pmodel 
is 
a tuple 
$\langle \mathcal{Q}, \Sigma, \Lambda, q_{init}, \transitionRel\rangle$,
in which $\mathcal{Q}$ is a non-empty set of states, 
$\Sigma$ is the non-empty finite set of \emph{input} symbols, 
$\Lambda$ is the non-empty finite set of \emph{output} symbols, 
$q_{init} \subseteq \mathcal{Q}$ is the initial state, 
and \transitionRel is the transition relation $\transitionRel\subseteq Q \times \Sigma \times \Lambda \times Q$. 
Given a transition $\langle q_1, \alpha, \gamma, q_2\rangle\in\transitionRel$, it signifies that 
if the protocol is in state $q_1\in\mathcal{Q}$ and receives the input symbol $\alpha$, then it will 
transition to state $q_2$ and will generate the output symbol $\gamma$. 
We consider this reception of input $\alpha$ and generation of output $\gamma$ as the 
\textit{observation} of the transition from $q_1$ to $q_2$.

The input and output alphabets $\Sigma$ and $\Lambda$ in our description 
are intentionally left to be abstract. In our context, $\Sigma$ can be viewed as the cross-product 
of all the input message types and predicates over the input message fields. Similarly, 
$\Lambda$ can be defined as the cross-product of all the output message types and predicates over them. 
For example, the input symbol \smcmdintegrityzero denotes the \smcmd message, with the predicate \zerointegrity 
denoting the message is not integrity protected, \shthree denoting the security header field value of 
the message is set to 3, and \eiaone denoting the EIA algorithm field is set to 1. 

\noindent\textbf{Trace.} A trace \trace is a  sequence of the form    
$[\alpha_1/\gamma_1, \alpha_2/\gamma_2, 
\ldots, \alpha_k/\gamma_k]$ where $\alpha_i\in\Sigma$ and $\gamma_i\in\Lambda$. 
In our context, \trace signifies a protocol execution in which 
the protocol implementation is fed the input symbols  
$[\alpha_1, \alpha_2, \ldots, \alpha_k]$, and it generates   
the output symbols $[\gamma_1, \gamma_2, \ldots, \gamma_k]$.
For instance, consider the trace \tracesup{\ast} in BLE \prelimexampleone. 
In \tracesup{\ast}, \scanreq has been first sent to the IUT that responds with \scanresp. Then,  \connectionreq is sent, to 
which the response is \nullaction (i.e., the IUT does not 
generate an output). After that, the response to \versionreqextllzero is 
\versionresp, and so on.
The length of trace \trace is denoted by $|\trace|$. The $i^\mathrm{th}$ 
element of \trace is denoted with \tracesub{i} (where $0\leq i < |\trace|$). 
``$\cdot$'' represents the concatenation of two traces. As an example, 
$[\alpha/\gamma] \cdot \tracesup{1}$ denotes a trace obtained by prepending the 
trace with a single observation $\alpha/\gamma$ to trace $\tracesup{1}$.

\begin{comment}
\noindent\textbf{Regular Expression(RE).}
A regular expression defines a regular language, which is a set of strings accepted by a deterministic finite automaton (DFA). The strings consist of symbols belonging to an alphabet $\mathcal{A}$. Detailed explanations regarding regular expressions and operations on them are provided in Appendix \ref{sec:appendix_re}. In our work, we only consider the Kleene star, union, and negation operation, and the union operation is limited to literals only. \textcolor{blue}{We limit ourselves only to these operations since these operators are sufficient to represent the violation of all the security properties we have tested.}
%\syed{Why do we consider only these operations? Why not other operators in general?}

We term each constituent regular expression, including any operation on it, as \textit{elements} of the regular expression. Each element can either be a \textit{literal element} having no kleene star operation or a \textit{kleene star element} consisting of a kleene star operation. As an example, consider the regular expression $\sigma_v = (.)^{*}\mathsf{a}_{pr}\neg({\mathsf{a}_{sc}})^{*}\mathsf{a}_{kf}$.
Here, there $\sigma_v$ has 4 elements, with $(.)^{*}$ and $\neg({\mathsf{a}_{sc}})^{*}$ being kleene star elements and ${a}_{pr}$ and $\mathsf{a}_{kf}$ being literal elements. We also denote $|\sigma|$ as the number of elements in $\sigma$, i.e., $|\sigma_v| = 4$.
\end{comment}
%\input{sections/3_protocol_testing_problem}
%\input{sections/4_motivation_new}
\newcommand{\fontGuidingFSMFig}{\fontsize{45}{45}\selectfont\bfseries}
\section{{Motivation of PROTEUS}}
\label{sec:motivation}

We first review the unique challenges of testing COTS wireless protocol implementations and then use a running example to motivate \system{}'s design choices.

\noindent\textbf{Closed-source implementations.} As COTS wireless 
devices tend to be closed-source, 
any testing approach relying on some level of access to the source code (white box or gray box) is inapplicable, warranting 
a black box testing approach which \system follows.

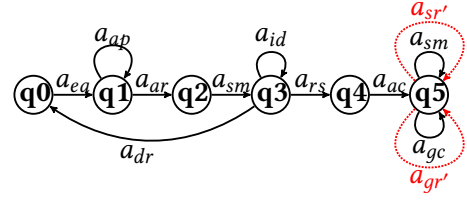
\begin{figure}[ht]
\centering
%\vspace{-0.2cm} 
\resizebox{0.8\linewidth}{!}{%
 \begin{tikzpicture}[>=stealth',bend angle=45,auto]

\tikzstyle{state}=[circle,thick,draw=black!100,fill=white!20,minimum size=1.2cm,text width=1.5cm,align=center,inner sep=0pt,line width=1mm]
\tikzstyle{edgeLabel}=[pos=0.5,text centered,text width=5cm,font={\sffamily\fontGuidingFSMFig}];  % Adjusted font size here
\tikzset{every loop/.style={min distance=1mm,looseness=3}}

\begin{scope}[->,node distance=4.2cm,yshift=0mm, xshift=-135mm, local bounding box=a_nodes]
% First net
\node [state, name=q0] {\fontGuidingFSMFig q0};
\node [state, name=q1, right of=q0, yshift=0mm] {\fontGuidingFSMFig q1};
\node [state, name=q2, right of=q1, yshift=0mm] {\fontGuidingFSMFig q2};
\node [state, name=q3, right of=q2, yshift=0mm] {\fontGuidingFSMFig q3};
\node [state, name=q4, right of=q3, yshift=0mm] {\fontGuidingFSMFig q4};
\node [state, name=q5, right of=q4, yshift=0mm] {\fontGuidingFSMFig q5};

\path[every node/.style={font=\sffamily\Large,inner sep=0pt}]
    (q0) edge[bend right=0, line width=1mm] node[edgeLabel,yshift=1mm,xshift=0mm] {\fontGuidingFSMFig$a_{ea}$} (q1)
    (q1) edge[bend right=0, line width=1mm] node[edgeLabel,yshift=1mm,xshift=0mm] {\fontGuidingFSMFig$a_{ar}$} (q2)

    % (q1) edge[red, loop above, out=120, in=60, looseness=5, yshift=20mm,  densely dashed, line width=1mm] node[edgeLabel, line width=1mm] {\fontGuidingFSMFig$a_{ac'}$} (q1)

    (q1) edge[loop below, out=125, in=55, looseness=6, line width=1mm] node[edgeLabel, yshift=10mm, xshift=0mm] {\fontGuidingFSMFig$a_{ap}$} (q1)

    (q2) edge[bend right=0, line width=1mm] node[edgeLabel,yshift=1mm,xshift=0mm] {\fontGuidingFSMFig$a_{sm}$} (q3)
    (q3) edge[bend right=0, line width=1mm] node[edgeLabel,yshift=1mm,xshift=0mm] {\fontGuidingFSMFig$a_{rs}$} (q4)
    (q3) edge[loop above, out=120, in=60, looseness=5, line width=1mm] node[edgeLabel, yshift=3mm, xshift=0mm, line width=1mm] {$a_{id}$} (q3)

    (q3) edge[bend right=-35, line width=1mm] node[edgeLabel,yshift=-2mm,xshift=-8mm] {\fontGuidingFSMFig$a_{dr}$} (q0)
    (q5) edge[loop above, out=120, in=60, looseness=5, line width=1mm] node[edgeLabel, yshift=1mm, xshift=0mm] {\fontGuidingFSMFig$a_{sm}$} (q5)
    (q5) edge[loop below, out=-120, in=-60, looseness=5, line width=1mm] node[edgeLabel, yshift=-1mm, xshift=0mm] {\fontGuidingFSMFig$a_{gc}$} (q5)

    % new add for mutation - ridwan
    (q5) edge[red, loop above, out=135, in=45, looseness=10, densely dashed, line width=1mm] node[edgeLabel, yshift=1mm, xshift=0mm] {\fontGuidingFSMFig$a_{sr'}$} (q5)
    
    (q5) edge[red, loop below, out=-135, in=-45, looseness=10, densely dashed, line width=1mm] node[edgeLabel, yshift=-1mm, xshift=0mm] {\fontGuidingFSMFig$a_{gr'}$} (q5)

    % (q1) edge[red, bend right=-40, densely dashed, line width=1mm] node[edgeLabel,yshift=2mm,xshift=-8mm] {\fontGuidingFSMFig$a_{ac'}$} (q5)

    (q4) edge[bend right=0, line width=1mm] node[edgeLabel,yshift=1mm,xshift=0mm] {\fontGuidingFSMFig$a_{ac}$} (q5);
\end{scope}
\end{tikzpicture}
}
% \vspace{-0.5cm}   
 \caption{A partial LTE protocol state machine used as guiding PSM for running example. The transition labels are presented in Table \ref{tab:ble_running_example_table}, 
 with red transitions indicating mutations.
 %Transitions marked red denote mutated transitions.
 }
\label{fig:ble_running_example}
\vspace{-0.3cm}   
\end{figure}

\noindent\textbf{Need for a faithful test oracle.} Crashing behavior can serve as a protocol-agnostic universal symptom for memory-related bugs. In contrast, semantic or logical bugs require a faithful test oracle to accurately adjudicate the correct behavior for a given test case. \system uses a set of security properties collected from RFCs/specifications 
%and validated manually 
as faithful oracles. If a property is violated, \system directly concludes that it is a logical vulnerability.  

\noindent\textbf{Protocol statefulness.} Statefulness of wireless protocols introduces the following challenges: 
(1) Analysis mechanisms must be aware of the underlying protocol state. (2) The protocol behavior depends on both the current state and the current protocol packet.
(3) Triggering a bug may require driving the target protocol implementation in a particular protocol state. This may require sending a long sequence of messages to the IUT (to reach the bug-triggering state) before we can inject the bug-triggering packet. (4) Due to 2 and 3, the universe of test cases is potentially infinite. A naive sampling of this infinite test case universe is unlikely to be effective for a given testing budget. \system, therefore, uses guidance from both the PSM and properties to efficiently sample meaningful test cases that will likely trigger a vulnerability. 
    
%Also, due to point (1) above, each test case must be executed when the protocol implementation is in the initial state. Otherwise, a previous test case may drive the protocol to an unknown state for which the test oracle does not know the correct behavior. Also, reproducing a buggy behavior in such a case would require storing all test cases executed before the bug-triggering test case. We must reset the device after executing each test case to avoid this. If a protocol does not have a known \emph{homing sequence}, one has to resort to a hard device reset (\,  e.g., rebooting a device), substantially increasing the amortized cost of executing a single test case.
\noindent\textbf{High cost of running a test case.} Because of protocol statefulness, any previous test case may drive the protocol to an unknown state for which the test oracle does not know the correct behavior. 
%\textcolor{red}{Reproducing a buggy behavior in such a case would require storing all test cases executed before the bug-triggering test case.} 
%Therefore, we must reset the device after executing each test case. 
If a protocol does not have a known \emph{homing sequence}{~\cite{stulman2009searching}}, one has to reset the device's 
state machine either using a sequence of OTA messages or reboot the device. Also, after sending each protocol message in a test case, the tester has to wait a certain time to check whether the message induces a device crash. The resets and device responsiveness checks substantially increase the amortized cost of executing a single test case.
%Thus, hard resets and IUT  responsiveness checks contribute to the high cost of running a test case. 
As an example, it takes up to $\sim\!\!1.5$ minutes in BLE and $\sim\!\!1$ minute in LTE to run a test case containing $\sim\!\!6\!-\!8$ messages. The high execution cost significantly slows the overall testing speed and allows testing only a limited number of test cases within a given time budget. Failing to explore more protocol behavior within a specific time results in fewer bug detections. 

\noindent\textbf{Testing budget awareness.} 
Traditional fuzzing approaches for general-purpose systems \emph{implicitly} consider a \emph{testing budget}. They try to accelerate the discovery of security vulnerabilities through different mechanisms such as enhancing code coverage~\cite{chen2018angora, lemieux2018fairfuzz, zhou2020zeror}, optimizing the search space of test messages ~\cite{feng2021snipuzz}, adopting efficient scheduling mechanisms ~\cite{lyu2019mopt}, increasing testing speed ~\cite{xu2017designing}, or improving discovery of execution paths ~\cite{gan2018collafl}. 
%Many other approaches focus on coverage improvement to optimize testing ~\cite{chen2018angora, lemieux2018fairfuzz, zhou2020zeror}. 
%These approaches, however, \emph{implicitly} consider a \emph{testing budget} and try to optimize their fuzzing approach to maximize the discovery of vulnerabilities. 
However, these approaches have no explicit mechanism to adapt their test case generation according to the available testing budget. In contrast, \system \emph{explicitly} respects the testing budget. 
%It prioritizes efficient and effective vulnerability discovery within a given budget by leveraging the guiding PSM and input properties and explicitly controlling the mutation amount and length of the test cases to generate meaningful test cases . 
It %prioritizes efficient and effective vulnerability discovery within a given budget by 
leverages a guiding PSM and input properties and explicitly controls the mutation amount and length of the test cases to generate meaningful test cases that can be executed within the available time budget.

\vspace{-0.2cm}
\subsection{Running Example}
To justify the approach taken by \system, we consider a running example 
using 4G LTE. 

\noindent\textbf{Guiding PSM.} 
The partial PSM used in this example, as shown in Figure~\ref{fig:ble_running_example}, is extracted from the LTE NAS layer protocol specification. 
Table \ref{tab:ble_running_example_table} explains the transition labels of the guiding PSM. 
%are explained in Table \ref{tab:ble_running_example_table}. 
Some errors are introduced intentionally (red transitions) in guiding PSM.
% \vspace{-1mm}

\begin{table}[t]
\centering
\caption{Transition labels (\ie, input and output symbols) in the guiding PSM.}
\vspace{-0.4cm}
\resizebox{\linewidth}{!}{
\begin{tabular}{|l|p{10cm}|}
\hline
\textbf{Symbol} & \textbf{Description}                                                                      \\ \hline
$a_{ea}$        & {\normalsize\textsc{enable\_s1 / attach\_request}\normalsize}                           \\ \hline
$a_{ar}$        & {\normalsize\textsc{authentication\_request / authentication\_response}\normalsize}     \\ \hline
$a_{sm}$        & {\normalsize\textsc{security\_mode\_command / security\_mode\_complete}\normalsize}     \\ \hline
$a_{id}$        & {\normalsize\textsc{identity\_request: integrity == 1 \& identity\_type == 1/ identity\_response}\normalsize}                 \\ \hline
% $a_{id'}$        & {\normalsize\textsc{identity\_request: integrity == 0 and identity\_type == 1/ identity\_response}\normalsize}                 \\ \hline
$a_{dr}$        & {\normalsize\textsc{detach\_request / detach\_accept}\normalsize}                 \\ \hline

$a_{rs}$        & {\normalsize\textsc{rrc\_security\_mode\_command / rrc\_security\_mode\_complete}\normalsize}                 \\ \hline
 
$a_{ac}$        & {\normalsize\textsc{attach\_accept / attach\_complete}\normalsize}                     \\ \hline
$a_{gc}$        & {\normalsize\textsc{guti\_reallocation\_command / guti\_reallocation\_complete}\normalsize}                        \\ \hline
% $a_{gr}$        & {\normalsize\textsc{guti\_reallocation\_command: replay == 1 / null\_action}\normalsize}         \\ \hline
$a_{gr'}$        & {\normalsize\textsc{guti\_reallocation\_command: replay == 1 / guti\_reallocation\_complete}\normalsize}         \\ \hline
$a_{sr'}$        & {\normalsize\textsc{security\_mode\_command: replay == 1 / security\_mode\_complete}\normalsize} 
\\ \hline
$a_{ap}$        & {\normalsize\textsc{attach\_accept: integrity == 0 \& cipher == 0 \& security\_ header \_type == 0/ null\_action}\normalsize}                     \\ \hline
$a_{ap'}$        & {\normalsize\textsc{attach\_accept: integrity == 0 \& cipher == 0 \& security\_ header \_type == 4/ attach\_complete}\normalsize}
\\ \hline
\end{tabular}
}
%\captionsetup{justification=centering}
\label{tab:ble_running_example_table}
\vspace{-0.3cm}
\end{table}

%\vspace{-6mm}
\noindent\textbf{Desirable property.}
The security property of interest in this example is the following. 
$\phi_g:$ ``\emph{After successfully completing the attach procedure,
a replayed GUTI Reallocation Command message should not be accepted.}''
This property prevents an attacker from performing linkability attacks by violating the freshness of a device's ephemeral identity (\ie, GUTI). Violating this property enables an adversary to launch an attack in which they send a previously captured \gutiral message intended for the victim to all devices in a cell. If the victim device 
violates the above property and is present in that cell, it will respond positively, whereas others will just 
respond with a rejection. In this way, the adversary can test the presence of a victim 
in a cell. 

\subsection{Benefit of Having PSM and Properties}
Before explaining the advantage of having access to both a guiding PSM and desirable security properties for generating meaningful test cases within a testing budget, we first explain why any approach having access to just one of these is unlikely to be as effective. 

%, with a running example,
%In the following discussion, we use the example discussed above to motivate the discovery of vulnerabilities 
%due to the violation of $\phi_g$. 

\noindent\textbf{Why is guiding PSM not enough?} Consider a PSM-guided testing approach that mutates a trace adhering to the protocol flow to generate a test case. However, this approach does not guarantee that the mutated test case will violate $\phi_g$; wasting a test case. As an example, consider a good protocol flow sampled from the PSM denoted as \textbf{S0} 
in Table \ref{tab:example_sequences_ble_runningexample}.
% \omar{Please add the good flow here}. 
Suppose we mutate \textbf{S0} by adding a mutation on the \idreq message to generate the test case \textbf{S1} 
in Table \ref{tab:example_sequences_ble_runningexample}.  
\textbf{S1} clearly does not violate the property as it does not even include the \gutiral message, let alone its replay. 
\begin{table}[t]
%\scriptsize
 \caption{Sequences (test traces) considered for the running example in Figure \ref{fig:ble_running_example}. %Explanation of the symbols are presented in Table \ref{tab:ble_running_example_table}.
    }
    \vspace{-0.3cm}
\centering
	\begin{tabular}{|c|l|}
            
		\hline
		\textbf{ID} & \textbf{Sequence}\\ \hline
        
        % \textbf{S1} & \textbf{P1} & \textit{\textsl{scan\_req/scan\_resp, pair\_req:sc==0/null\_action, version\_req:ext\_ll==0/null\_action, scan\_req/scan\_resp, key\_exchange/null\_action}} \\ \hline
        \textbf{S0} & $a_{ea}, a_{ar}, a_{sm}, a_{id}$ \\ \hline
        
        \textbf{S1} & $a_{ea}, a_{ar}, a_{sm}, \pmb{a_{id'}}$ \\ \hline
        
        \textbf{S2} & $a_{ea}, a_{ar}, a_{sm}, \pmb{a_{dr}}, a_{ac}, a_{gc}, a_{gr'}$ \\ \hline
        %\textbf{S4} &\textbf{P3} & \textit{\textsl{scan\_req/scan\_resp, con\_req/null\_action, version\_req:ext\_ll==0/version\_resp, pair\_req:sc==1/pair\_resp, \textbf{dh\_check:valid==0/null\_action}, key\_exchange/null\_action}} \\ \hline
        %\textbf{S5} &\textbf{P3} & \textit{\textsl{scan\_req/scan\_resp, con\_req/null\_action, version\_req:ext\_ll==0/version\_resp, pair\_req:sc==1/pair\_resp, \textbf{dh\_check:valid==0/null\_action}, \textbf{dh\_check:valid==0/null\_action}, key\_exchange/null\_action}} \\ \hline
        
        \textbf{S3} & $a_{ea}, a_{ar}, a_{sm}, a_{rs}, a_{ac}, a_{gc}, \pmb{a_{sr'}}, \pmb{a_{gr'}}$ \\ \hline

%        \textbf{S4} & \tracesfour \\ \hline

	\end{tabular}
	% \vspace*{-0.1cm}
        %\captionsetup{justification=centering}
       
    % \\ \ishtiaq{TODO: which property, impact} 
 %        \begin{flushleft}
		
	% \end{flushleft}
	\label{tab:example_sequences_ble_runningexample}
	\vspace*{-0.3cm}
\end{table}

\noindent\textbf{Why are properties not enough?} 
Access to properties, however, ensures that such blatantly 
vacuous test cases like \textbf{S1} are not generated. Considering the property $\phi_g$, one can automatically 
generate \emph{test skeletons} that are guaranteed to violate it. One such test skeleton or template 
(expressed as a regular expression for ease of exposition) is 
$\sigma_g = (.)^{*}\mathsf{a}_{ar}(.)^{*}\mathsf{a}_{sm}(.)^{*}\mathsf{a}_{ac}(.)^{*}\mathsf{a}_{gc}(.)^{*}\mathsf{a}_{gr'}$ where 
wildcard characters signify that they can be replaced with 0 or more occurrences of any input symbols, and still result in a $\phi_g$-violating 
test. $\sigma_g$ captures the necessary messages to violate $\phi_g$, with wildcard characters allowing other protocol messages. Suppose we instantiate $\sigma_g$ by replacing wildcard characters with arbitrary concrete symbols. Without any knowledge of a typical good protocol flow, we may obtain instantiated test case \textbf{S2} (see Table \ref{tab:example_sequences_ble_runningexample}) which violates $\phi_g$. However, any practical protocol implementation would most likely reject this trace since $a_{dr}$ will discard the NAS security context and the subsequent messages of \textbf{S2} will be dropped. Also, a successful RRC security context $a_{rs}$ is not performed, preventing the completion of the attach procedure. 
Such blind instantiations of wildcard characters to generate a test case will likely result in meaningless test cases that the IUT will reject, wasting test cases.

\noindent\textbf{Advantage of \system{}' approach.}
\system takes advantage of the strengths of both PSM-guided and property-guided approaches. 
The main insight \system uses is that after a trace skeleton like $\sigma_g$ is created, it does 
not blindly instantiate the wildcard characters and instead relies on the PSM for guidance on instantiating these free choices. When instantiating these open choices, it does not necessarily follow the PSM exactly but also includes controlled deviations. For example, using the PSM, \system would instantiate $\sigma_g$ and generate a test case \textbf{S3}, which not only violates 
$\phi_g$ but also have a higher chance of being accepted by a buggy implementation as it closely follows a good protocol flow. 
\section{Design Overview and Challenges}
\label{sec:overview_and_challenges}

\begin{figure}[t]
    \centering
    %\vspace{-0.2cm}
    \includegraphics[width=0.9\linewidth]{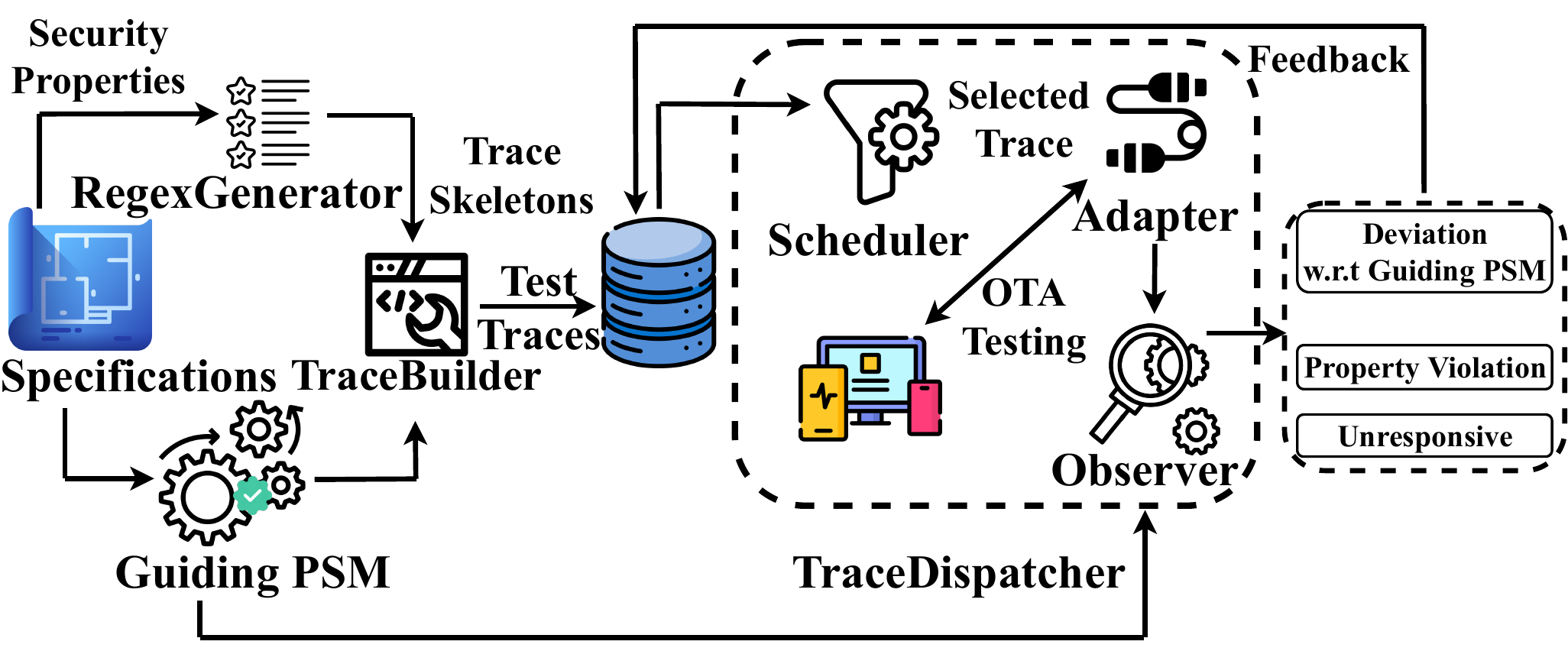}
    %\vspace{-0.2cm}
     \vspace{-0.4cm}
    \caption{Overview of \system.
    % \ishtiaq{TODO: will update later after everything is finalized.}
    }
    \label{fig:system_overview}
\vspace{-0.4cm}      
\end{figure}

\subsection{Proteus Overview} 
\label{subsec:overview}

Given a set of desired properties $\Phi = \{\phi_1, \phi_2, \ldots, \phi_n\}$, 
a (potentially, standard-prescribed) guiding PSM \pmodel of a protocol $P$, 
a testing budget $\beta$, and an implementation \pimp of the protocol $P$ under test (i.e., IUT), 
\system aims to identify execution traces $\tracesup{c}$ of \pimp such that it 
falsifies one or more desired properties $\phi_i\in\Phi$ while receiving 
guidance from \pmodel and $\Phi$ in time proportional to $\beta$. $\beta$ 
is of the form $\langle \lambda, \mu\rangle$ in which $\lambda$ denotes the maximum 
length budget (\ie, the number of input symbols in a test case) and 
$\mu$ represents the mutation budget (\ie, the maximum number of places a test case 
can deviate from a good protocol flow). Figure~\ref{fig:system_overview} presents the high-level overview of \system.  
The full pseudocode of \system{}'s test generation is presented in Algorithm \ref{algo:proteus_approach} in Appendix.
Conceptually, for each property $\phi_i \in \Phi$, \system carries out the following steps. 

\textbf{(1) Test skeleton generation.} \system uses its \ltlparser
to generate a test skeleton denoted as a regular expression (RE) for $\phi_i$. 
% (lines 3-5 in Algorithm 1 shown in our extended paper ~\cite{proteus_extended}). 
% \ishtiaq{please check if this is okay}
This test skeleton is an abstract execution/trace \tracesup{a} of the protocol under test guaranteed to violate $\phi_i$. Some positions of \tracesup{a} have specific input 
symbols, whereas others have wildcard characters. One can instantiate wildcards with input symbols and still violate $\phi_i$. Based on $\beta$, \system will generate multiple such test skeletons. 
    
\textbf{(2) Instantiating test skeletons.} In this step, \system instantiates the abstract 
    trace \tracesup{a}, especially in unrestricted positions (represented with wildcard 
    characters) with specific input symbols under the guidance of \protomachine 
    while respecting the budget $\beta$. 
    % (lines 6-9 of the Algorithm shown in the full version of the paper ~\cite{proteus_extended}) \ishtiaq{check}
     Based on $\beta$, \system will generate multiple instantiated test cases \tracesup{c}, containing input symbols in all positions with no wildcard characters, and store them in \textit{TestTraceSet}. Guidance from \protomachine is crucial to ensure that any generated instantiated trace is indeed meaningful.
% such test skeletons.  
   
    \textbf{(3) Dispatching test cases.} \system then takes each instantiated \tracesup{c}, schedules it, and then prepares OTA protocol packets for each input symbol. It then sends 
    the resulting concrete protocol packets contained in this concretized test case OTA to \pimp. 
    
    \textbf{(4) Vulnerability detection and reporting.} If \pimp accepts the concretized test case executed OTA, 
    then it signifies a vulnerability. In such a case, the test case is stored, and a vulnerability is reported.  
    
%\vspace{-2mm}
\subsection{Challenges and Insights}
\label{subsec:design_challenges}

Realizing \system{}' approach requires tackling the following challenges. We address the challenges with the outlined novel insights.

\noindent\ding{111} \textbf{Challenge C1: Synthesizing test skeletons from properties.} 
\system takes the desired properties as past linear temporal logic (PLTL) formulae. 
We, therefore, first require to generate test skeletons that are guaranteed to violate a property. Also, we need a succinct representation for test skeletons. 
% Given the desired properties as past linear temporal logic (PLTL) formulae as input to \system,  
% we must generate test skeletons that are guaranteed to violate a property represented. Also, we need a succinct representation for test skeletons. 

\noindent\textbf{Insight I1.}  
We use regular expressions (REs) as our succinct intermediate representation of 
a property-violating test skeleton. RE is not only expressive enough to capture 
the temporal ordering in a PLTL formula but also can represent logical dependencies. This leads to the following research question: 
\emph{how does one generate \tracesup{a} in regular expression format from a PLTL formula automatically?} 

{
To answer this, we observe that a PLTL expression contains logical (e.g., AND operator) and temporal (e.g., SINCE operator) operators. Logical operators of the PLTL formula express constraints over its operands for any \tracesup{a} expressing the violation of the PLTL expression $\phi$. In contrast, temporal operators express constraints over their operands in a range of positions in \tracesup{a}. An abstract syntax tree (AST) effectively expresses the relation between an operator and its operands, and its leaves are propositions required to obtain the input symbol or wildcard characters in \tracesup{a}. \system parses the AST of the PLTL formula and leverages the PLTL operator semantics to non-deterministically generate a trace skeleton \tracesup{a}.
}

\noindent\ding{111}\textbf{ Challenge C2: Instantiating abstract test skeletons using guiding PSM.}
The next challenge is to instantiate the abstract test skeleton $\tracesup{a}$ to a test case $\tracesup{c}$ 
while taking guidance from the PSM. {We assume that the guiding PSM satisfies all security properties. In contrast, property-violating traces will likely be present when implementations 
subtly deviate from the standard. As such, we opt to perform \emph{mutations} on the guiding PSM and generate traces to instantiate the abstract test skeleton.
}

However, if we perform too many mutations on the guiding PSM $\mathcal{M}$, 
we will generate traces that substantially deviate from the standard, which are unlikely to 
trigger any vulnerabilities. In contrast, if we perform too few mutations, we may miss some vulnerabilities that require more mutations to identify (e.g., generating a test trace that detects the GUTI reallocation 
replay attack in LTE requires two mutations). Thus, we need to vary the amount of mutation $\mu$ within a range of values.

Similarly, generating arbitrarily long traces negatively impacts testing. Long traces likely repeat the same states and transitions, making it unlikely to uncover new vulnerabilities. Even worse, excessively long traces 
takes significantly more time to test OTA, ultimately wasting our testing budget. 
In contrast, we need our test traces to be at least as long as the number of literals in the trace skeleton \tracesup{a} to generate a trace that satisfies \tracesup{a}. 
We also require a slightly longer trace 
% than the number of literals in \tracesup{a}
to visit extra states and uncover potential property violations. Thus, similar to mutations, 
we need to vary the length of the generated traces $\lambda$ within a range of values.

Finally, for a given test skeleton \tracesup{a}, the challenge is to design an automatic approach to generate instantiated traces \tracesup{c} that would (1) satisfy the abstract test skeleton \tracesup{a}, (2) align with the guiding PSM \pimp, (3) ensure that each generated instantiated trace \tracesup{c} require at most $\mu$ mutations on $\mathcal{M}$, and (4) maintain a maximum length of $\lambda$. For example, consider the guiding PSM in {Figure \ref{fig:ble_running_example}} and the property $\phi_g$ provided in Section \ref{sec:motivation}. A trace skeleton that violates $\phi_g$ is  $\sigma_g$, and \textbf{S3} (shown in Table \ref{tab:example_sequences_ble_runningexample}) is a test trace that satisfies $\sigma_g$. \textbf{S3} aligns with $\mathcal{M}$ within 2 mutations (i.e., it would satisfy any requirement of $\mu >= 2$). The mutations are marked bold in Table \ref{tab:example_sequences_ble_runningexample}. Also, \textbf{S3} is of length 8 and thus satisfies any length requirement of $\lambda >= 8$.

\noindent\textbf{Insight I2.} 
We observe that we can solve the problem of generating test cases 
described above by combining \textit{overlapping subproblems} 
originating from destination states $q_n$ ($q_n \in \mathcal{M}$) of the outgoing 
transitions of a particular state $q_c$. The solution of the subproblems will produce traces by 
using mutations wherever necessary to satisfy some suffix of \tracesup{a} starting from state $q_n$. 
%\syed{Check the previous sentence.} 
By prepending an observation (depending on \tracesup{a}) to these traces,
%\syed{What is ``suitable observation''? Automated vs. manual?}
we can generate traces originating from $q_c$ that satisfy \tracesup{a} 
within the specified mutation and length budget. For example, in Figure 
\ref{fig:ble_running_example} if we have that trace $\trace_k = a_{ar}a_{sm}a_{sm}a_{rs}a_{ac}a_{gc}a_{sm'}a_{gr'}$ 
that satisfies $\sigma_g$ from state $q_1$, and append observation $a_{ea}$ (obtained from the transition 
from $q_0$ to $q_1$) to this trace,  we will obtain trace \textbf{S3} shown in 
Table \ref{tab:example_sequences_ble_runningexample}. \textbf{S3} satisfies $\sigma_g$ from $q_0$ and requires the same amount of mutation but has one length more than $\trace_k$. 
Also, consider two subproblems involving two transitions with the same observation and destination state but different source states. We want to satisfy the same trace skeleton with the same budget. 
Then, any trace obtained from the first subproblem will also be a solution for the second subproblem. Thus, the subproblems exhibit \textit{overlapping} characteristics and \system adopts a dynamic programming-based solution to generate test cases.

\noindent\ding{111} \textbf{Challenge C3}: \textbf{Arbitrary mutations miss logical vulnerabilities.}  
As discussed in challenge \textbf{C2}, one must perform mutations on the PSM to increase the chance of triggering security property violations. To mutate a PSM, we must select a transition and alter it, i.e., its input, output, or destination state. At any protocol state, randomly selecting a transition to mutate is less likely to violate the given property since it would not be property-driven. Similarly, even after selecting a transition to mutate, randomly altering the input message or output message of the transition at the bit/byte level, like traditional fuzzers~\cite{afl, afl++}, would also be inefficient since it does not consider the semantic meaning of the message fields (i.e., byte offset and boundary of each message field), the given security property, and the current protocol state. Although grammar-guided fuzzers ~\cite{Boofuzz} offer semantic meaning-aware mutations, they do not consider the property being tested and the current state of the protocol while selecting a message to mutate. Such mutation schemes are less likely to generate mutated messages that violate a given security property. Therefore, an effective mutation scheme should (i) consider the security property being tested while selecting a transition and (ii) consider the semantic meaning of a message, the given property, and the current protocol state while mutating a selected message {(i.e., have a clever choice to fill the wildcard characters of the test skeleton \tracesup{a}).}

\noindent\textbf{Insight I3.}
To address challenge \textbf{C3}, we consider selecting a transition to mutate $\mathcal{M}$ with a goal to generate traces satisfying test skeleton \tracesup{a}. Such a mutation scheme is \textit{property driven} since we aim to satisfy \tracesup{a}, which signifies the violation of its corresponding security property. Also, while mutating any transition in $\mathcal{M}$, we consider mutating a transition's observation or destination state. Mutating the observation may uncover improper handling of prohibited messages. In contrast, mutating the destination state may uncover certain bypass attacks since these attacks represent skipping certain intermediate states to reach a secure state in protocol registration/authentication procedures. Moreover, to perform a mutation over the input message, we consider performing operations on it that are semantic aware (e.g., understanding field boundaries of any message field), state aware (e.g., setting field values within or outside a defined range according to the current protocol state) and also property driven (e.g., creating a plaintext version of a message to test at a security context established state, considering the property that plaintext messages should be dropped at such a state). Consequently, our mutation scheme is more likely to uncover a violation of the given security property than other existing works.

\noindent\ding{111} \textbf{Challenge C4:} 
\textbf{Arbitrarily scheduling the properties and test traces to test OTA will lead to inefficiency.} 
Once we generate the set of test traces for all given properties, one can randomly schedule test traces to execute over-the-air (OTA). 
%Our objective is to identify the maximum number of property violations. 
However, this approach is inefficient as it may fail to uncover vulnerabilities or adequately explore the search space within the given testing budget.

\noindent\textbf{Insight I4.}
To address challenge \textbf{C4}, \system adopts an efficient scheduling mechanism to uncover vulnerabilities faster. In each testing iteration, \system first selects a property $\phi$ from the property set $\Phi$ to test and then a trace $\pi^{c}$ to test $\phi$. \system prioritizes scheduling the properties whose generated traces cover more states in the guiding PSM, increasing the likelihood of vulnerability detection. After selecting a property $\phi$, \system selects a trace $\pi^{c}$, prioritizing based on the frequency of use of $\pi^{c}$, the number of already identified deviations $\pi^{c}$ covers, and the number of instances of $\pi^{c}$ rendered the target unresponsive. To avoid testing redundant traces, \system{} does not test any further traces of a property whose violation has already been detected.

% If a property violation is detected, \system does not test any other traces from that property, avoiding testing redundant traces and enabling \system to focus on testing other properties. 

% \begin{table}[t]
% \begin{tabular}{|c|c|c|}
% \hline
% \textbf{Testcase ID} & \textbf{\begin{tabular}[c]{@{}c@{}}Properties \\ Violated\end{tabular}} & \textbf{\begin{tabular}[c]{@{}c@{}}Selection\\ Order\end{tabular}} \\ \hline
% $\pi^{1}$                   & $\phi_1$, $\phi_2$                                                                  & 2                                                                  \\ \hline
% $\pi^{2}$                   & $\phi_2$                                                                      & 3                                                                  \\ \hline
% $\pi^{3}$                   & $\phi_1$, $\phi_2$, $\phi_3$                                                              & 1                                                                  \\ \hline
% \end{tabular}
% \caption{Simple scheduling example for insight \textbf{I4}}
% \label{tab:scheduling_example}
% \end{table}

\begin{table}[]
\caption{Simple scheduling example for insight \textbf{I4}.}
\vspace{-0.3cm}
\resizebox{.8\linewidth}{!}{
\begin{tabular}{|c|c|>\centering m{2.5cm}|>\centering m{2.5cm}|c|}
\hline
\textbf{\begin{tabular}[c]{@{}c@{}}Test case \\ ID\end{tabular}} & \textbf{\begin{tabular}[c]{@{}c@{}}Properties \\ Violated\end{tabular}} & \textbf{\begin{tabular}[c]{@{}c@{}}Average  States\\ Covered\end{tabular}} & \textbf{\begin{tabular}[c]{@{}c@{}}No. of Deviations\\ Covered\end{tabular}} & \textbf{\begin{tabular}[c]{@{}c@{}}Selection\\ Order\end{tabular}} \\ \hline
$\pi^{1}$                                                       & $\phi_1$                                                                & 5                                                                  & 2                                                                                 & 1                                                                  \\ \hline
$\pi^{2}$                                                       & $\phi_1$                                                                & 5                                                                  & 1                                                                                 & x                                                                \\ \hline
$\pi^{3}$                                                       & $\phi_1$                                                                & 5                                                                  & 0                                                                                 & x                                                                \\ \hline
$\pi^{4}$                                                       & $\phi_2$                                                                & 3                                                                  & 2                                                                                 & 2                                                                \\ \hline
\end{tabular}
}
\vspace{-0.4cm}
\label{tab:scheduling_example}
\end{table}
\newcommand{\fontRegexFig}{\fontsize{30}{30}\selectfont}
As a simple example of our scheduling scheme, consider four traces, their corresponding property, the average number of states covered by the traces associated with the property, the number of already identified deviations covered by the trace, and their selection order in Table \ref{tab:scheduling_example}. Since traces associated with $\phi_1$ cover more states on average than $\phi_{2}$, \system first selects $\phi_1$ to test. Among the three traces associated with $\phi_1$, suppose all other factors are the same, but trace $\pi^{1}$ covers more transitions where \system observed a deviation from $\mathcal{M}$ in previous iterations. \system then selects $\pi^{1}$ to test OTA. Now suppose $\pi^{1}$ identified a violation of property $\phi_1$. Then, \system discards testing all traces associated with $\phi_1$ (i.e., $\pi^{2}$ and $\pi^{3}$), and in the next iteration selects $\pi^{4}$ to test.
%\vspace{-0.8cm}
%\input{sections/6_design_challenges}
%\input{sections/5_high_level_approach}
\section{\ltlparser: Constructing Test Skeletons From Security Properties}
\label{sec:regex_generator}

\system first constructs test skeleton(s) for each property $\phi \in \Phi$ expressed in PLTL formula and uses the skeletons represented in regular expressions to generate test cases. To automatically construct test skeletons, we have developed 
\ltlparser that takes a PLTL (past linear temporal logic) formula as input and produces REs violating the PLTL formula as output. 
To construct such an RE $\sigma_i$, \ltlparser leverages the PLTL formula's Abstract Syntax Tree (AST) and traverses it in pre-order. 
Nodes in this AST correspond to PLTL operators (e.g.,  Since $S$, Yesterday $Y$), while leaves represent observations  $(\alpha/\gamma)$. For instance, Figure \ref{fig:regex-generator} presents the AST for the PLTL formula $\phi = $ \mybox{$a_{sm}\!\implies \! !a_{id'}~S~a_{dr}$}.

To create a violating regular expression for a PLTL formula $\phi$, at each internal (non-leaf) node $n_i$ during AST traversal, \textsf{RegexGenerator} needs to satisfy or negate the sub-formula $\phi^{n_i}$ rooted at $n_i$ based on the semantic meaning of the operators involved in $\phi^{n_i}$. 
For this, \ltlparser recursively satisfies or negates the expression at $n_i$'s child nodes according to the semantic meaning of the operator at $n_i$. While traversing a leaf node $n_l$ of the AST, \ltlparser places a literal or a kleene star element on $\sigma$ according to the requirement (satisfaction or negation) at $n_l$. 
For example, if the operator is a ``\emph{since}'' as in  \mybox{$!a_{id'}~S~a_{dr}$} in Figure~\ref{fig:regex-generator}, \ltlparser attempts to first avoid satisfying the right subtree $a_{dr}$ by placing $\neg(a_{dr})^{*}$, and then violate the left subtree by placing $a_{id'}$ after $\neg(a_{dr})^{*}$. Thus it will find $\neg(a_{dr})^{*}a_{id'}$ violating the ``\emph{since}'' operator. Again, for the ``\emph{implies}'' operator, \ltlparser attempts to first satisfy the left subtree (obtaining RE ${(.)}^{*}a_{sm}$) and then violate the right subtree (obtaining RE $\neg(a_{dr})^{*}a_{id'}$). 
Finally, to construct an RE representing the violation of the given PLTL formula, the operator/operand at the root node of the AST must be negated. 
The final RE $\sigma_v$ violating $\phi$ will be \mybox{${(.)}^{*}a_{sm}\neg(a_{dr})^{*}a_{id'}$}.
\ltlparser generates multiple different violating expressions. It also checks if a previously generated RE already covers the expression. If so, it discards the new RE.

\begin{figure}[t]
\centering
\resizebox{0.1\textheight}{!}{%
\begin{tikzpicture}[>=Stealth, bend angle=45, auto]
\tikzstyle{state}=[circle, thick, draw=black!100, fill=white!20, minimum size=1.2cm, text width=1.5cm, align=center, inner sep=0pt]

\tikzstyle{rect}=[rectangle, draw, text width=8.5cm, text centered, minimum height=1.5cm]
\tikzstyle{rect_wide}=[rectangle, draw, text width=8.5cm, text centered, minimum height=1.5cm]

\begin{scope}[->, node distance=1.2cm, yshift=0mm, xshift=-135mm, local bounding box=a_nodes]
\node [state, name=a1] {\fontRegexFig H};
\node [state, name=a2, below of=a1, yshift=-10mm] {\fontRegexFig $\Rightarrow$};
\node [state, name=a3, below left of=a2, yshift=-8mm, xshift=-10mm] {\fontRegexFig $a_{sm}$};
\node [state, name=a4, below right of=a2, yshift=-8mm, xshift=10mm] {\fontRegexFig S};

\node [state, name=a5, below left of=a4, yshift=-9mm, xshift=-10mm] {\fontRegexFig \textbf{!}};
\node [state, name=a6, below right of=a4, yshift=-9mm, xshift=10mm] {\fontRegexFig $a_{dr}$};
\node [state, name=a7, below of=a5, yshift=-12mm]  {\fontRegexFig $a_{id'}$};
\node [rect_wide, name=rect1, above of=a1, yshift=16mm]  {\fontRegexFig $H({a}_{sm} \Rightarrow \textbf{!} {a}_{id'} \ S \ a_{dr})
$};

\node [rect, name=rect2, below of=a7, yshift=-15mm]  {\fontRegexFig $\left(.\right)^* a_{sm} \neg \left(a_{dr}\right)^* a_{id'}
$};

\path
    (a1) edge[bend right=0, line width=0.5mm] (a2)
    (a2) edge[bend right=0, line width=0.5mm] (a3)
    (a2) edge[bend right=0, line width=0.5mm] (a4)
    (a4) edge[bend right=0, line width=0.5mm] (a5)
    (a4) edge[bend right=0, line width=0.5mm] (a6)
    (a5) edge[bend right=0, line width=0.5mm] (a7)
    (rect1) edge[bend right=0, line width=1.2mm] (a1)
    (a7) edge[bend right=0, line width=1.2mm] (rect2)
    ;

\end{scope}
\end{tikzpicture}
}
\vspace{-0.3cm}
\caption{Example AST of a PLTL formula.}
\label{fig:regex-generator}
\vspace{-0.3cm}
\end{figure}
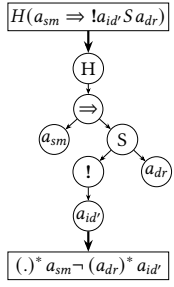

\newcommand{\fontGuidingFSMLabel}{\fontsize{55}{55}\selectfont\bfseries}

\section{\dpalgo: Generating Test Cases from Test Skeletons and Guiding PSM}
\label{sec:tracebuilder}
Given a test skeleton \tracesup{a} representing violations of a security property $\phi \in \Phi$, \system uses a guiding PSM $\mathcal{M}$ and mutates $\mathcal{M}$ to efficiently generate meaningful test cases \tracesup{c} with the shape of \tracesup{a}. 

{Note that the guiding PSM used by Proteus can be abstract compared to PSMs extracted from target device implementations. As such, they are substantially smaller than PSMs learned from commercial devices (\S\ref{sec:experiments}). One can also use PSMs directly obtained from RFCs (e.g.,~\cite{rfc9293}) as a guiding PSM. Since \system does not require detailed PSMs, it alleviates the prohibitively high time required to learn a detailed PSM from target implementations.} 
We now discuss our PSM mutation strategies and   
then present our test case generation mechanism.  

%\vspace{-2mm}
\subsection{Mutating a PSM}
\label{sec:tracebuilder:mutation}

As discussed in challenge \textbf{C2} in \S\ref{sec:overview_and_challenges}, since the guiding PSM is assumed to satisfy all security properties, we need to mutate the guiding PSM $\mathcal{M}$ to generate traces satisfying a test skeleton \tracesup{a}, which signifies the violation of a security property. \system selects a transition $\mathcal{R}(q_c, \alpha, \gamma, q_n)$ to mutate while generating instantiated traces satisfying a test skeleton \tracesup{a} (described in \S\ref{sec:tracebuilder:trace_generation}). 
For mutations, \system essentially performs two types of operations on a selected transition $\mathcal{R}$, as discussed below. 

\noindent\ding{111} \textbf{Mutation kind M1: Mutating the observation of a transition.} 
In this case, \system{} alters the observation $\alpha/\gamma$ of $\mathcal{R}$,  i.e., alters either the input $\alpha$ or the output $\gamma$ or both based on the input and output in trace skeleton \tracesup{a} when the execution reaches at $q_c$.

At any state $q_c$, if the trace skeleton \tracesup{a} to satisfy requires input $\alpha'$ and output $\gamma'$ but the guiding PSM $\mathcal{M}$ does not have $\alpha'/\gamma'$at $q_c$, to ensure the generated trace conforms with \tracesup{a}, we mutate the transition with $\alpha'$ at $q_c$ in $\mathcal{M}$ to obtain input $\alpha'$ and output $\gamma'$. For example, in Figure 1, suppose at state $q_5$ if we require output \footnotesize\textsc{GUTI\_ reallocation \_complete}\normalsize\xspace for input
\footnotesize\textsc{GUTI\_ reallocation \_command: Replay ==1}\normalsize\xspace according to the test skeleton. However, since it is not prescribed at state $q_5$, we mutate the transition with observation $a_{gc}$ (with input \footnotesize\textsc{GUTI\_ reallocation \_command}\normalsize\xspace) to obtain input \footnotesize\textsc{GUTI\_ reallocation \_command: Replay == 1}\normalsize\xspace and output \footnotesize\textsc{GUTI\_ reallocation \_complete}\normalsize\xspace.
This mutation strategy is required to generate traces satisfying $\tracesup{a}$.

\newcommand{\fontMutOpTable}{\fontsize{8}{8}\selectfont}

\def\arraystretch{1}

% \vspace{-10mm}
\begin{table}[t!]
\caption{List of possible semantic operations performed on an input message for mutation type M1. \footnotesize\textsc{Hop}\normalsize\xspace is a 5-bit field in \connectionreq message in BLE whose value is defined to be between 5 to 16 according to BLE specifications. \attachacc and \smcmd are messages in 4G LTE.}
\vspace{-0.3cm}
\resizebox{\linewidth}{!}{
%\scriptsize
\centering
	\renewcommand{\arraystretch}{0.9}
	\fontMutOpTable
 
\begin{tabular}{
  |>{\centering\arraybackslash}m{1.25cm}|m{4.5cm}|m{4.2cm}|
}
\hline
\textbf{Operation} & \textbf{Description} & \textbf{Example}                                                      \\ \hline
% \textbf{OP1}       & Adopt a provided mutated version of the message  &            \footnotesize\textsc{guti\_ reallocation\_ command: Replay == 1}\normalsize\xspace for matching $a_{gr'}$ in $\pi^{a}$          \\ \hline
\textbf{OP1}       & Change value of a field to a defined value in range  &            \footnotesize\textsc{connection\_ request: Hop == 5}\normalsize\xspace          \\ \hline
\textbf{OP2}       & Change value of a field to prohibited value/value outside of defined range &  \footnotesize\textsc{connection\_ request: Hop == 20}\normalsize\xspace\\ \hline
\textbf{OP3}       & Change value of a field to one of its boundary values (i.e., setting all bits of the field to 0 or 1)     &             \footnotesize\textsc{connection\_ request: Hop == 31}\normalsize\xspace                \\ \hline
\textbf{OP4}       & Send the plaintext version of any integrity protected and/or ciphered message &  \footnotesize\textsc{attach\_accept: Integrity == 0 \& Cipher == 0}\normalsize\xspace\\ \hline
\textbf{OP5}       & Combination of applying OP1, OP2, OP3 and OP4 multiple times    &   \footnotesize\textsc{attach\_accept: Integrity == 0 \& Cipher == 0 \& Security\_Header\_Type == 15}\normalsize\xspace                                \\ \hline
\textbf{OP6}       & Replay a previously captured version of the message       &   \footnotesize\textsc{security\_mode\_command: Replay == 1}\normalsize\xspace                               \\ \hline
\end{tabular}
}
%\vspace{-0.3cm}
\label{tab:mutation_ops_m1}
\end{table}

On the other hand, if \tracesup{a} has a wildcard character at state $q_c$, \system{} instantiates the wildcard character with a mutated version of an input message $\alpha$ to detect any deviation from the PSM that can lead to a security property violation. 
To increase the likelihood of a mutated message being accepted by a practical protocol implementation, \system{} considers the semantic meaning of the input message. It performs one of the six semantic operations. We summarize these operations with examples in Table \ref{tab:mutation_ops_m1}. These operations consider the semantics of the message fields, their defined values and ranges according to the protocol specification, and overall message semantics (e.g., plaintext or replayed version). As an example, in Figure \ref{fig:mutation_kind_example}(i), consider a mutation of the transition with observation $a_{ap}$ (with input \footnotesize\textsc{attach\_ accept: integrity == 0 \& cipher == 0 \& security\_ header \_ type == 0}\normalsize\xspace, marked blue) at state $q_1$. If we perform \textbf{OP2} operation and change the value of the \footnotesize\textsc{security\_ header \_ type}\normalsize\xspace  field to a prohibited value 4, we obtain the mutated message $a_{ap}'$ with input \footnotesize\textsc{attach\_ accept: integrity == 0 \& cipher == 0 \& security\_ header \_ type == 4}\normalsize\xspace. If the target accepts this mutated message, we obtain \footnotesize\textsc{attach\_complete}\normalsize\xspace as a response, which deviates from the guiding PSM.
%\vspace{-0.2cm}

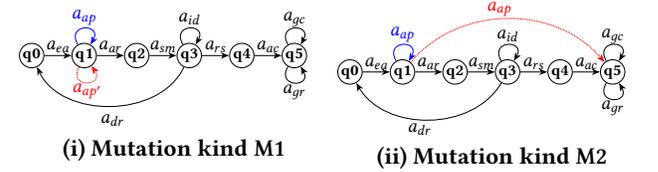
\begin{figure}[htbp]
\centering
% Begin first minipage for M1
\begin{minipage}{0.49\linewidth}
\centering
\resizebox{\linewidth}{!}{
\begin{tikzpicture}[>=stealth',bend angle=45,auto]

\tikzstyle{state}=[circle,thick,draw=black!100,fill=white!20,minimum size=2cm,text width=2cm,align=center,inner sep=0pt,line width=1mm]
\tikzstyle{edgeLabel}=[pos=0.5,text centered,text width=7cm,font={\sffamily\fontGuidingFSMLabel}];  % Adjusted font size here
\tikzset{every loop/.style={min distance=1mm,looseness=3}}

\begin{scope}[->,node distance=5.2cm,yshift=0mm, xshift=-135mm, local bounding box=a_nodes, every edge/.append style={-{Stealth[scale=1.5]}}]
% First net
\node [state, name=q0] {\fontGuidingFSMFig q0};
\node [state, name=q1, right of=q0, yshift=0mm] {\fontGuidingFSMFig q1};
\node [state, name=q2, right of=q1, yshift=0mm] {\fontGuidingFSMFig q2};
\node [state, name=q3, right of=q2, yshift=0mm] {\fontGuidingFSMFig q3};
\node [state, name=q4, right of=q3, yshift=0mm] {\fontGuidingFSMFig q4};
\node [state, name=q5, right of=q4, yshift=0mm] {\fontGuidingFSMFig q5};

\path[every node/.style={font=\sffamily\Large,inner sep=0pt}]
    (q0) edge[bend right=0, line width=1mm] node[edgeLabel,yshift=2.5mm,xshift=0mm] {\fontGuidingFSMLabel$a_{ea}$} (q1)
    (q1) edge[bend right=0, line width=1mm] node[edgeLabel,yshift=2.5mm,xshift=0mm] {\fontGuidingFSMLabel$a_{ar}$} (q2)

    % (q1) edge[red, loop above, out=120, in=60, looseness=5, yshift=20mm,  densely dashed, line width=1mm] node[edgeLabel, line width=1mm] {\fontGuidingFSMFig$a_{ac'}$} (q1)

    (q1) edge[blue, loop above, out=120, in=60, looseness=5, line width=1mm] node[edgeLabel, yshift=3mm, xshift=0mm] {\fontGuidingFSMLabel$a_{ap}$} (q1)

    (q2) edge[bend right=0, line width=1mm] node[edgeLabel,yshift=2.5mm,xshift=0mm] {\fontGuidingFSMLabel$a_{sm}$} (q3)
    (q3) edge[bend right=0, line width=1mm] node[edgeLabel,yshift=2.5mm,xshift=0mm] {\fontGuidingFSMLabel$a_{rs}$} (q4)
    (q3) edge[loop above, out=120, in=60, looseness=5, line width=1mm] node[edgeLabel, yshift=3mm, xshift=0mm, line width=1mm] {$a_{id}$} (q3)

    (q3) edge[bend right=-65, line width=1mm] node[edgeLabel,yshift=-8mm,xshift=2mm] {\fontGuidingFSMLabel$a_{dr}$} (q0)
    (q5) edge[loop above, out=120, in=60, looseness=5, line width=1mm] node[edgeLabel, yshift=1mm, xshift=0mm] {\fontGuidingFSMLabel$a_{gc}$} (q5)
    (q5) edge[loop below, out=-120, in=-60, looseness=5, line width=1mm] node[edgeLabel, yshift=-1mm, xshift=0mm] {\fontGuidingFSMLabel$a_{gr}$} (q5)

    % new add for mutation - ridwan
    (q1) edge[red, loop below, out=-115, in=-55, looseness=5, densely dashed, line width=1mm] node[edgeLabel, yshift=-1mm, xshift=0mm] {\fontGuidingFSMLabel$a_{ap'}$} (q1)

    % (q1) edge[red, bend right=-40, densely dashed, line width=1mm] node[edgeLabel,yshift=2mm,xshift=-8mm] {\fontGuidingFSMFig$a_{ac'}$} (q5)

    (q4) edge[bend right=0, line width=1mm] node[edgeLabel,yshift=2.5mm,xshift=0mm] {\fontGuidingFSMLabel$a_{ac}$} (q5);
\end{scope}
\end{tikzpicture}
}
\textbf{(i) Mutation kind $\mathbf{M1}$}
\end{minipage}
\begin{minipage}{0.49\linewidth}
\centering
\resizebox{\linewidth}{!}{
\begin{tikzpicture}[>=stealth',bend angle=45,auto]
    
    \tikzstyle{state}=[circle,thick,draw=black!100,fill=white!20,minimum size=2cm,text width=2cm,align=center,inner sep=0pt,line width=1mm]
    \tikzstyle{edgeLabel}=[pos=0.5,text centered,text width=7cm,font={\sffamily\fontGuidingFSMLabel}];  % Adjusted font size here
    \tikzset{every loop/.style={min distance=1mm,looseness=3}}
    
    \begin{scope}[->,node distance=5.2cm,yshift=0mm, xshift=-135mm, local bounding box=a_nodes, every edge/.append style={-{Stealth[scale=1.5]}}]
    % First net
    \node [state, name=q0] {\fontGuidingFSMFig q0};
    \node [state, name=q1, right of=q0, yshift=0mm] {\fontGuidingFSMFig q1};
    \node [state, name=q2, right of=q1, yshift=0mm] {\fontGuidingFSMFig q2};
    \node [state, name=q3, right of=q2, yshift=0mm] {\fontGuidingFSMFig q3};
    \node [state, name=q4, right of=q3, yshift=0mm] {\fontGuidingFSMFig q4};
    \node [state, name=q5, right of=q4, yshift=0mm] {\fontGuidingFSMFig q5};
    
    \path[every node/.style={font=\sffamily\Large,inner sep=0pt}]
        (q0) edge[bend right=0, line width=1mm] node[edgeLabel,yshift=2.5mm,xshift=0mm] {\fontGuidingFSMLabel$a_{ea}$} (q1)
        (q1) edge[bend right=0, line width=1mm] node[edgeLabel,yshift=2.5mm,xshift=0mm] {\fontGuidingFSMLabel$a_{ar}$} (q2)
    
        % (q1) edge[red, loop above, out=120, in=60, looseness=5, yshift=20mm,  densely dashed, line width=1mm] node[edgeLabel, line width=1mm] {\fontGuidingFSMFig$a_{ac'}$} (q1)
    
        (q1) edge[blue, loop above, out=120, in=60, looseness=5,line width=1mm] node[edgeLabel, yshift=3mm, xshift=0mm] {\fontGuidingFSMLabel$a_{ap}$} (q1)

        (q2) edge[bend right=0, line width=1mm] node[edgeLabel,yshift=2.5mm,xshift=0mm] {\fontGuidingFSMLabel$a_{sm}$} (q3)
        (q3) edge[bend right=0, line width=1mm] node[edgeLabel,yshift=2.5mm,xshift=0mm] {\fontGuidingFSMLabel$a_{rs}$} (q4)
        (q3) edge[loop above, out=120, in=60, looseness=5, line width=1mm] node[edgeLabel, yshift=3mm, xshift=0mm, line width=1mm] {$a_{id}$} (q3)
    
        (q3) edge[bend right=-60, line width=1mm] node[edgeLabel,yshift=-3mm,xshift=-9mm] {\fontGuidingFSMLabel$a_{dr}$} (q0)
        (q5) edge[loop above, out=120, in=60, looseness=5, line width=1mm] node[edgeLabel, yshift=1mm, xshift=0mm] {\fontGuidingFSMLabel$a_{gc}$} (q5)
        (q5) edge[loop below, out=-120, in=-60, looseness=5, line width=1mm] node[edgeLabel, yshift=-1mm, xshift=0mm] {\fontGuidingFSMLabel$a_{gr}$} (q5)
    
        % new add for mutation - ridwan
        % (q5) edge[red, loop below, out=-135, in=-45, looseness=10, densely dashed, line width=1mm] node[edgeLabel, yshift=-1mm, xshift=0mm] {\fontGuidingFSMFig$a_{gr'}$} (q5)
    
        (q1) edge[red, bend right=-50, densely dashed, line width=1mm] node[edgeLabel,yshift=2mm,xshift=-8mm] {\fontGuidingFSMLabel$a_{ap}$} (q5)

        (q4) edge[bend right=0, line width=1mm] node[edgeLabel,yshift=2.5mm,xshift=0mm] {\fontGuidingFSMLabel$a_{ac}$} (q5);
    \end{scope}
    \end{tikzpicture}
}
\textbf{(ii) Mutation kind $\mathbf{M2}$}
\end{minipage}
\caption{Two kinds of mutations M1 and M2. The transition labels are presented in Table \ref{tab:ble_running_example_table}.}
\label{fig:mutation_kind_example}
\end{figure}
%\vspace{-0.1cm}

\noindent\ding{111} \textbf{Mutation kind M2: Mutating the destination state of a transition.} 
In this case, \system{} alters the destination state $q_n$ of a selected transition $\mathcal{R}$ $(q_c, \alpha, \gamma, q_n)$. The goal of this kind of mutation is to test whether the input message $\alpha$ at state $q_c$ leads to a state $q_m$ different than what is expected by the guiding PSM (in this case $q_n$), which in turn can lead to an observable deviation or security property violation. This kind of mutation enables \system to identify certain bypass attacks, e.g., detect whether any intermediate step can be bypassed to reach a secure state. To illustrate, in Figure \ref{fig:mutation_kind_example}(ii), suppose we perform mutation over the transition with observation $a_{ac'}$ at state $q_1$ (marked blue). We can perform a mutation over the next state $q_2$ of the transition by changing the next state of the transition to $q_5$. This mutation would generate an instantiated trace that tests whether the target incorrectly reaches state $q_5$ if a plaintext \attachacc is fed at $q_1$.

%\vspace{-0.3cm}
\subsection{Instantiated Trace Generation}\label{sec:tracebuilder:trace_generation}

Given a test skeleton \tracesup{a} representing violation of a property $\phi \in \Phi$, a guiding PSM $\mathcal{M}$, a mutation budget $\mu$ and a length budget $\lambda$, we leverage insight \textbf{I2} and solve the problem of generating instantiated traces by combining the solutions of overlapping subproblems. For this, we formulate a dynamic programming problem as follows.

% \begin{tcolorbox}
$\mathbf{\mathcal{G}(\mathcal{M}, \tracesup{a}, q_c, \mu, \lambda)}$: \emph{At any state $\mathsf{q}_c \in \mathcal{Q}$ of a guiding PSM $\mathcal{M}$, we want to craft a set of instantiated traces $\mathcal{T}$, where each instantiated trace (i) satisfies the test skeleton \tracesup{a}, (ii) is the outcome of a maximum of $\mu$ mutations applied on any trace in the guiding PSM $\mathcal{M}$ and (iii) is within a maximum length $\lambda$.}     
% \end{tcolorbox}

%We use $\mathcal{G}(\mathcal{M}, \tracesup{a}, q_c, \mu, \lambda)$ to denote the problem which we can solve by leveraging the solution to its \emph{overlapping subproblems}. 
Consider we are at state $q_i$ of our guiding PSM $\mathcal{M}$ and we need to satisfy the $j^{th}$ observation $l_j$ = $(\alpha_j/\gamma_j)$ of \tracesup{a}, i.e., the test skeleton up to input symbol $l_{j-1}$ has been satisfied and the next wildcard character in \tracesup{a} is $l_k$. Also, at $q_i$, consider the remaining mutation budget is $\mu_{i}$, and the remaining length budget is $\lambda_i$. For this problem, we consider the following four cases, formulating subproblems and prepending suitable observations.

\textbf{Case I.} If there is a transition $\mathcal{R}(q_i, \alpha, \gamma, q_m)$ at state $q_i$ whose observation satisfies $l_j$, \dpalgo can leverage solutions from the destination state $q_m$ that satisfies \tracesup{a} from observation $l_{j+1}$, with mutation budget $\mu_{i}$ and length budget $\lambda_i\!-\!1$. \dpalgo prepends $l_j$ to the instantiated traces being generated from the subproblem. Since it places an observation ($l_j$) in the instantiated trace, it uses length budget $\lambda_i\!-\!1$ to formulate the subproblem.

\textbf{Case II.} If there is no transition at $q_i$ satisfying $l_j$, \dpalgo can consume a mutation to mutate the transition for $\alpha_j$, the input message of observation $l_j$. \dpalgo places observation $l_j$ in the instantiated trace using mutation kind \textbf{M1}. Again, \dpalgo leverages traces from the destination state $q_m$ that satisfies \tracesup{a} from observation $l_{j+1}$, with length budget $\lambda_i\!-\!1$ but using mutation budget $\mu_i\!-\!1$ since it consumed a mutation to place $l_j$. \dpalgo prepends $l_j$ to the obtained instantiated traces from the subproblem.

\textbf{Case III.} If there is a transition $\mathcal{R}(q_i, \alpha, \gamma, q_m)$ at state $q_i$ whose observation $(\alpha/\gamma)$ satisfies wildcard character element $l_k$, \textsf{TraceBuilder} leverages solutions from the destination state $q_m$ that satisfies \tracesup{a} from observation $l_{j}$ with mutation budget $\mu_{i}$ and length budget $\lambda_i\!-\!1$. It prepends observation $(\alpha/\gamma)$ to the obtained instantiated traces from the subproblem.

\textbf{Case IV.} Finally, for any transition $\mathcal{R}(q_i, \alpha, \gamma, q_m)$ at state $q_i$, \dpalgo can mutate the transition's observation and place a mutated observation $(\alpha/\gamma)_m$ using mutation kind \textbf{M1}. \dpalgo can leverage solutions from the destination state $q_m$ that satisfies \tracesup{a} from observation $l_{j}$ with length budget $\lambda_i\!-\!1$ and also mutation budget $\mu_i\!-\!1$ since we perform a mutation. It prepends the mutated observation $(\alpha/\gamma)_m$ to the obtained instantiated traces from the subproblem. Note that we only place a \textit{mutation marker} in this case. This marker would be resolved later by \dispatcher (\S\ref{sec:trace_dispatcher}). Note that we can generate multiple traces to test OTA by placing various mutated messages in place of the mutation marker. 

Furthermore, for each case, \dpalgo may also consider mutating the destination state of the selected transition (mutation kind \textbf{M2}). In that case, \dpalgo mutates the destination state to another state $q_{mut}$ and formulates the subproblem from state $q_{mut}$ with mutation budget $\mu\!-\! 1$. \dpalgo terminates if either \tracesup{a} is satisfied or runs out of mutation or length budget.
%\vspace{-0.2cm}

\section{\dispatcher: Test Execution and Flaw Detection}
\label{sec:trace_dispatcher}

After \dpalgo generates instantiated traces associated with each property in its property set $\Phi$, the \dispatcher component of \system iteratively selects a property $\phi$ and then an instantiated test trace \tracesup{c} generated from $\phi$ to test OTA. \dispatcher runs for $t$ iterations, where $t$ is proportional to $\beta$. \dispatcher consists of three components: a \textit{scheduler}, an \textit{adapter} and an \textit{observer}. 
%We now describe the three components in detail.

\noindent\textbf{Scheduler.} In each testing iteration, the \textit{scheduler} of \textsf{TraceDispatcher} first chooses a property $\phi \in \Phi$ to test using weighted random sampling. The scheduler determines the weight of each property by determining the average number of distinct states in the guiding PSM covered by all instantiated traces of $\phi$. This scheme favors properties whose instantiated traces cover more states within the guiding PSM and implicitly prioritizes traces more likely to violate a security property within the input property set $\Phi$. 

{Once a property $\phi$ is selected, the scheduler chooses an instantiated trace \tracesup{c} to test from all traces associated with $\phi$. Note that an instantiated trace may or may not have mutation markers (case IV in \S\ref{sec:tracebuilder:trace_generation}). The scheduler prioritizes selecting instantiated traces with mutation markers since they test mutated input messages, which are more likely to trigger property violations.} Furthermore, if the scheduler chooses an instantiated trace with a mutation marker, it first prioritizes scheduling traces that mutate messages that were not mutated in previous iterations. Again, among the traces that have messages not mutated previously, the \textit{scheduler} selects an instantiated trace based on three factors: (i) the frequency of selection ($f$) of \tracesup{c}; (ii) the number of deviations covered by \tracesup{c} ($d$); (iii) the number of instances where \tracesup{c} rendered the IUT \pimp unresponsive ($u$). The scheduler selects the instantiated trace with the minimum score ($p = f - d + u$), randomly choosing one in case of a tie. The scheduler prioritizes instantiated traces that trigger known deviations since they can lead \pimp to an inconsistent state where \system is more likely to find property violations. Also, the scheduler is less inclined to traces that rendered \pimp unresponsive since we cannot find further property violations from an unresponsive \pimp. The scheduler randomly performs one of the six operations defined in Table \ref{tab:mutation_ops_m1} to mutate a message with a mutation marker and obtain a mutated input message.

\noindent\textbf{Adapter.} The adapter takes an instantiated trace selected by the scheduler to test OTA. It translates the abstract input symbols into concrete messages and sends them to \pimp. Also, it records the response from \pimp, converts it back into an abstract symbol, and sends the response sequence back to the observer.

\noindent\textbf{Observer.} The observer analyzes the response from \pimp to an instantiated trace and detects whether it violates any security properties. If there is any deviation from the guiding PSM, the observer automatically checks for property violation by checking the trace against the violating test skeletons (represented as REs) generated from the security properties. Additionally, to determine unresponsiveness, the \textit{observer} identifies the final protocol state $q_f$ according to the guiding PSM for each test trace. It then executes a message $\alpha_f$ OTA expected to elicit a valid output message $\gamma_f$. If \pimp does not respond, it is deemed unresponsive. 

%violates any security property by matching the trace with the violating test skeletons (represented as REs) generated from the security properties.

%\system uses the input security properties, collected from RFCs/specifications and validated manually, as faithful test-oracles. As such, if a property is violated, \ the system directly concludes that it is a logical vulnerability.  

%It also stores relevant information (instances of deviation, unresponsiveness, any message that was mutated) that the scheduler uses to schedule test cases.
%\vspace{-0.3cm}
%\input{section/fsm_and_properties}
% Also, for LTE we consider 112 security properties, and for BLE we consider 27 security properties
%\vspace{1cm}
\section{Experiments}
\label{sec:experiments}
We evaluate \system with the 4G LTE's NAS and RRC layer protocols (e.g., mobility management procedures) and BLE's SMP and Link Layer protocols ~\cite{ble_spec} %. %We evaluate \system 
based on the following research questions:

\begin{itemize}[noitemsep,topsep=0pt,leftmargin=0.3cm]
    \item \textbf{RQ1.} How effective is \system in finding novel and known issues in LTE and BLE implementations (\S\ref{sec:identified_issues})?

    \item \textbf{RQ2.} How effective and efficient \system is compared to existing works (\S\ref{sec:eval_comparison_existing_works})? 
    
    \item \textbf{RQ3.} How does \system perform with respect to generating test cases (\S\ref{sec:eval_performance_proteus})? 

\end{itemize}

\subsection{Experiment Setup For Testing}
We combine LTE's NAS and RRC layers' protocols and construct a single guiding PSM consisting of 7 states with 86 transitions in total. For BLE, our guiding PSM has 15 states with 244 transitions in total. Compared to PSMs learned from implementations ~\cite{dikeue,blediff}, the guiding PSMs in \system{} are substantially smaller. For example, an average-sized PSM learned from Pixel3A ~\cite{dikeue} reportedly had 21 states and 548 transitions, which is three times larger than
\system{} guiding PSM for 4G LTE.

For LTE, we set up a base station using srsRAN and a USRP B210, and a core network using srsEPC. We run them using an Intel i7-8665U processor with 16GB RAM. For BLE, we use nRF52840 acting as a central to test peripheral devices. We have tested 11 LTE and 12 BLE devices. Note that we updated all target devices with the latest patches before performing testing. The details of our tested devices (including SoC model, vendor, and baseband), and the identified issues in each device are provided in Tables \ref{tab:device_info_lte} and \ref{tab:device_info_BLE} in Appendix. 

\section{Identified Issues}
\label{sec:identified_issues}

% \NumLTEProperties properties
%\NumBLEProperties properties on
To answer \textbf{RQ1},
%we evaluate \system on \NumLTEDevice LTE and \NumBLEDevice BLE devices. We 
we have tested each device with 3000 OTA queries using \system.
Our evaluation reveals that \system identifies \NumLTEVulInstanece and \NumBLEVulInstanece vulnerabilities
in LTE and BLE, respectively, with \NumLTENew unique new issues in LTE and \NumBLENew unique new issues in BLE implementations. 
%We obtained five new CVEs, two bug bounties, and 7 acknowledgments from various vendors such as Google, Samsung, Qualcomm, and Microchip. 
The identified issues resulted in five new CVEs, two bug bounties, and 9 acknowledgments from various vendors such as Google, Samsung, Qualcomm, and Microchip.

Tables \ref{tab:lte-attacks} and \ref{tab:ble-attacks} 
summarize the identified issues for 4G LTE and BLE, respectively. 
The tables show the number of distinct devices where each issue was identified, the impact of each attack, and any new CVE/acknowledgment obtained by \system{} for each attack. We discuss in detail some of our identified issues in 4G LTE and BLE in \S\ref{sec:new-issue-lte} and \S\ref{sec:new-issue-ble}, respectively.

% \newcommand*{\priority}[1]{\begin{tikzpicture}[scale=0.12]%
% 		\draw (0,0) circle (1);
% 		\fill[fill opacity=1,fill=black] (0,0) -- (90:1) arc (90:90-#1*3.6:1) -- cycle;
% \end{tikzpicture}}

% \newcommand{\newFinding}{\priority{100}}
% \newcommand{\halfNewFinding}{\priority{50}}
% \newcommand{\oldFinding}{\priority{0}}

% \begin{comment}

\newcommand{\fontLTEVulTable}{\fontsize{8}{8}\selectfont}

\def\arraystretch{1}
\begin{table*}[t!]
\caption[list=off]{
        Vulnerabilities identified by \system for COTS LTE devices. 
        NAS-SC: NAS security context establishment, AS-SC: AS security context establishment. 
        % Category 
        \categoryone: known vulnerability found on devices previously confirmed to have the attack,  \categorytwo: known vulnerability found on new device not previously reported to have the vulnerability, \categorythree: previously unknown vulnerability. 
        % $^*$: The device is known to have the vulnerability.
        E-Exploitable, I-Interoperability, O-Other issue.
        }
\vspace*{-0.3cm}
\resizebox{\linewidth}{!}{
%\scriptsize
\centering
	\renewcommand{\arraystretch}{1}
	\fontLTEVulTable
	\begin{tabular}{
  |>{\centering\arraybackslash}m{0.65cm}|m{10.5cm}|
   >{\centering\arraybackslash}m{1.1cm}|
   >{\centering\arraybackslash}m{1.1cm}|
   >{\centering\arraybackslash}m{3cm}|
   m{5cm}|
}
		\hline
		\centering \textbf{Issue} & \textbf{Description}  & \textbf{Category} & \textbf{\#Vuln. Instance} & \textbf{Impact} &  \textbf{Newly Obtained CVEs/ Acknowledgements}  \\ 
		\hline
        \textbf{L-E1} & Accepts plaintext \idreq \fontLTEVulTable After NAS-SC \cite{dikeue, sherlock} & \categorytwo & 3 & Location Tracking &  Marked as duplicate by Samsung \footnote{We reported the issue to the vendor, they replied the issue was already reported independently for the device before we reported. However, no previous work detected the issue on the same device.}  \\ \hline
        \textbf{L-E2} & Accepts plaintext \authreq \fontLTEVulTable after NAS-SC \cite{dikeue, sherlock} & \categorytwo  &  4 & Location Tracking, DoS  &  Marked as duplicate by Samsung  \\ \hline
        \textbf{L-E3} & Accepts replayed \smcmd after NAS-SC \fontLTEVulTable \cite{dikeue, sherlock} &  \categoryone \categorytwo  &  4 & Location Tracking  & Marked as duplicate by Unisoc \\ \hline
        \textbf{L-E4} & Accepts replayed \gutiral after attach procedure \fontLTEVulTable \cite{dikeue}  &  \categorytwo  &  2 & Location Tracking  &  Marked as duplicate by Unisoc  \\ \hline
        \textbf{L-E5} & Accepts \rrcsmcmd \fontLTEVulTable with EIA0 IE after NAS-SC \cite{doltest,dikeue, rupprecht2016putting} & \categorytwo & 1 & Security Bypass &  High Severity CVE from Google  \\ \hline
        \textbf{L-E6} & \rrcsmcmd  \fontLTEVulTable with EIA0 IE after NAS-SC causes unresponsiveness \cite{dikeue}  & \categoryone \categorytwo  &  4 & DoS & - \\ \hline
        \textbf{L-E7} & Accepts plaintext \countercheck \fontLTEVulTable before AS-SC \cite{doltest} & \categorytwo  & 4 & Fingerprinting &  Marked as duplicate by Unisoc\\ \hline
        \textbf{L-E8} & \authreq \fontLTEVulTable with separation bit 0 cause further security mode procedure failure & \categorythree  & 4 & DoS &  Medium Severity CVE from Qualcomm  \\ \hline
        \textbf{L-E9} & Respond to \authreq \fontLTEVulTable with header 3 with \smrej   & \categorythree & 1 & Fingerprinting  &  Acknowledged by Huawei \\ \hline
        \textbf{L-O1} & Respond to \idreqtmsi \fontLTEVulTable with  \idrespimsi before NAS-SC  & \categorythree & 4 & Fingerprinting  & - \\ \hline

	\end{tabular}
 }
	
        %\captionsetup{justification=centering}
        
        \vspace*{-0.1cm}
	\label{tab:lte-attacks}
\end{table*}

\subsection{Identified Issues in LTE}
\label{sec:new-issue-lte}

%\subsubsection{Attacker Model}
\noindent \textbf{Attacker model.}
Similar to prior works~\cite{doltest, ltefuzz, lteinspector}, we assume that a Dolev-Yao attacker~\cite{dolev1983security} knows the victim's Cell Radio Network Temporary Identity (C-RNTI) using the victim's phone number~\cite{rupprecht2019breaking, jover2016ltesecurityprotocolexploits, hussain2019privacy}
and then send malformed messages to the victim device using a fake base station (FBS)~\cite{hussain2019insecure, yang2019overshadow}. {For attacks \textbf{L-E3} and \textbf{L-E4}, the attacker requires an adversary in addition to a FBS to capture and replay messages.}
The attacker can also use a Machine-in-the-Middle (MitM) relay~\cite{lteinspector,rupprecht2019breaking} to exploit the vulnerabilities, as MitM relays are more powerful than FBS.

\begin{figure}[ht]
    \centering
    \begin{minipage}{0.49\linewidth}
        \centering
        \includegraphics[width=\linewidth]{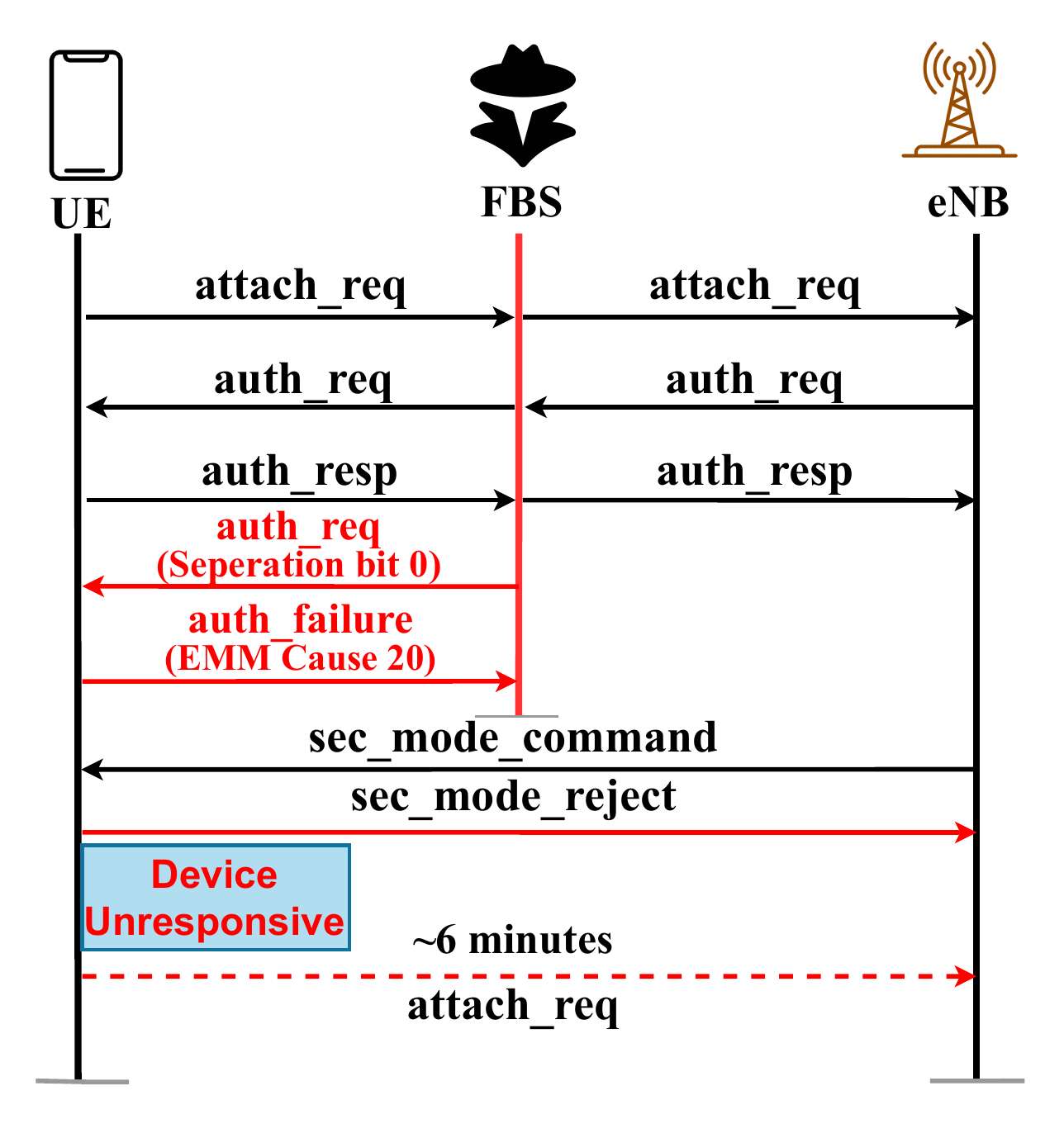}
        \vspace{-0.5cm}
        \subcaption{After authentication.}

    \end{minipage}
    \hfill
    \begin{minipage}{0.49\linewidth}
        \centering
            \includegraphics[width=\linewidth]{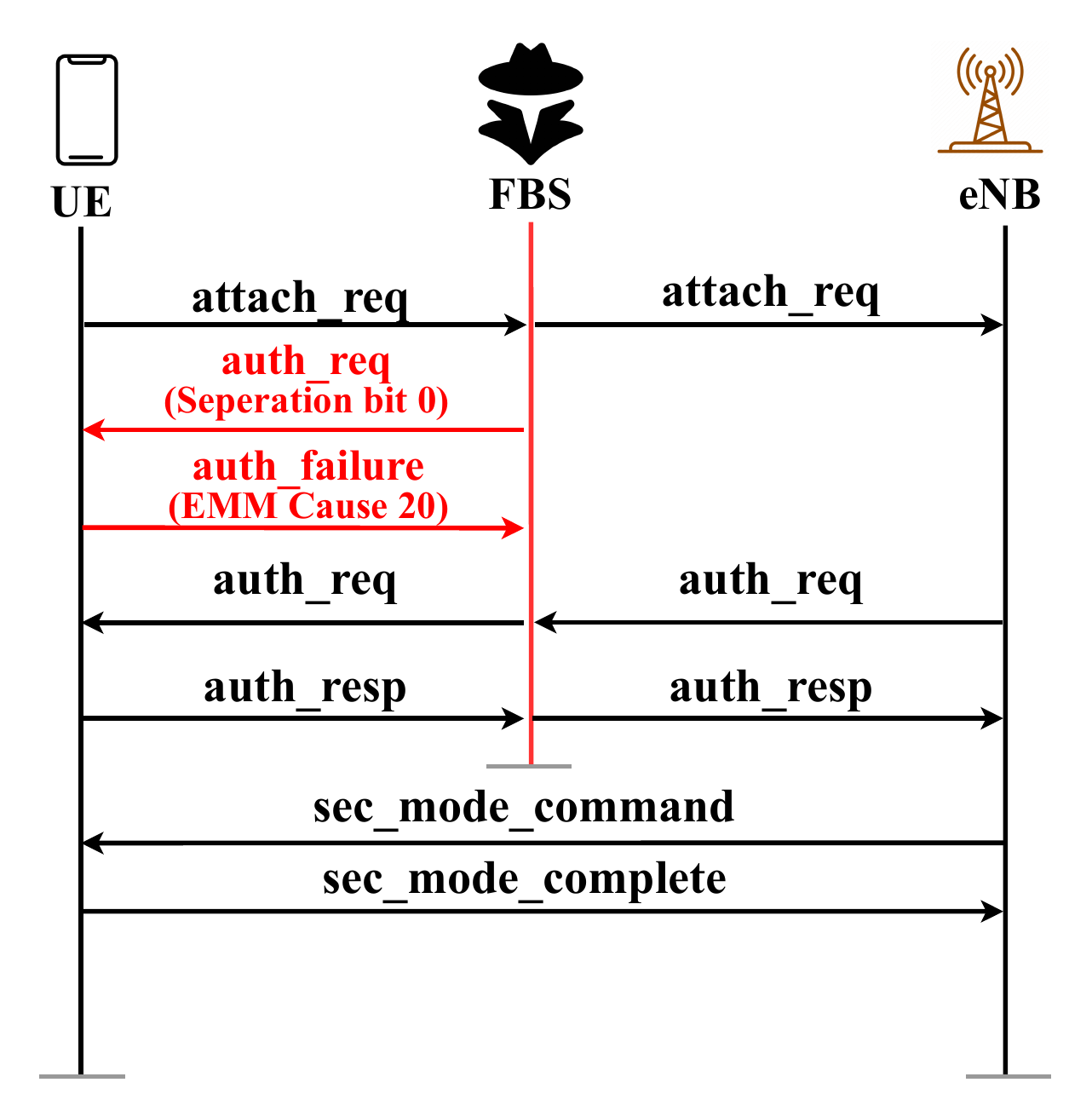}
        \vspace{-0.4cm}
        \subcaption{Before authentication.}
    
    \end{minipage}
    \vspace{-0.3cm}
    \caption{\authreq with separation bit 0 causing further security mode procedure failure.}
    \label{fig:issue_LE8}
    \vspace{-0.2cm}
\end{figure}

\noindent{\textbf{L-E8: \authreq with separation bit 0 causes further security mode procedure failure.}}
According to the 4G NAS specifications (TS 24.301, clause 5.4.2.6), a UE should send an \authfail message with EMM cause \# 26 
upon receiving an \authreq message with ``separation bit'' set to 0.
However, after successful authentication, affected devices respond with an \authfail message with EMM cause \#20 in response to such a message. The device then moves to ``unauthenticated'' state, leading to a subsequent failed security mode control. Figure~\ref{fig:issue_LE8}(a) shows the attack steps where the attacker sends such a malicious message to the victim UE using an FBS. Note that if the attacker attempts to send a malformed \authreq message before a successful authentication, the attack does not succeed (see Figure \ref{fig:issue_LE8}(b)). This signifies the stateful nature of the vulnerability that the existing works~\cite{doltest,ltefuzz} cannot detect.

\noindent\textbf{Impact.} 
Upon receiving the malformed message, Nexus 6P and Samsung A71 remained unresponsive for over 6 minutes and then attempted to reconnect. The attacker can repeat the attack steps, leading to a prolonged denial-of-service (DoS) attack on the victim user. Qualcomm identified the issue as medium severity and assigned a CVE. Note that the DoS attacks identified by \system are different in nature from spectrum jamming~\cite{pirayesh2022jamming, lichtman2016lte}, which are easily detectable, expensive, and unreliable. In contrast, our identified DoS attacks exploit logical vulnerabilities and allow targeting a single device with only one software-defined radio (SDR) and cannot be thwarted by jamming-specific defenses~\cite{jover2014enhancing}. 

\noindent{\textbf{L-E9: Respond to \authreq with header 3 with \smrej.}
} The affected devices incorrectly interpret a malformed \authreq with security header 3 before a successful authentication as a \smcmd, and respond with \smrej. However, in this scenario, the TS 24.301 ~\cite{4g_nas} clause 7.5.1 suggests returning a \emmstatus message with cause \#96 `invalid mandatory information'.

\noindent\textbf{Impact.} 
Since a correctly implemented device would drop this packet, an attacker can fingerprint the victim device up to the baseband manufacturer level and launch attacks by combining other known vulnerabilities at the baseband-chipset level. Huawei acknowledged this vulnerability as low severity.

\noindent\textbf{L-E5: Accepts \rrcsmcmd \fontLTEVulTable with EIA0 IE.} 
The affected devices respond with \rrcsmcmplt to the plaintext \rrcsmcmd with EIA0 message. This flaw leads to a remote privilege escalation attack with no additional execution privileges required. The attacker can bypass the RRC layer security activation by exploiting this vulnerability. The victim UE then accepts all plaintext RRC messages without integrity protection. \system{} detected this vulnerability in Pixel7. Google assigned a high-severity CVE to it. Although Rupprecht et al.~\cite{rupprecht2016putting} identified a similar issue in older devices, \system{} detected \textbf{L-E5} in new devices not previously reported to have the vulnerability. 
%\vspace{-0.4cm}

\subsection{Identified Issues in BLE}
\label{sec:new-issue-ble} 

\newcommand{\fontBLEVulTable}{\fontsize{8}{8}\selectfont}

\def\arraystretch{1}

% \vspace{-10mm}
\begin{table*}[t]
%\scriptsize
\caption{Vulnerabilities identified by \system for COTS BLE devices. All interpretations are the same as for Table \ref{tab:lte-attacks}. 
        }
\vspace*{-0.2cm}
\resizebox{\linewidth}{!}{
\centering
	\renewcommand{\arraystretch}{0.85}
	\fontBLEVulTable

\begin{tabular}{
  |>{\centering\arraybackslash}m{0.8cm}|m{11cm}|
   >{\centering\arraybackslash}m{1.1cm}|
   >{\centering\arraybackslash}m{1.1cm}|
   >{\centering\arraybackslash}m{2.1cm}|
   m{5cm}|
}
		\hline
		\centering \textbf{Issue} & \textbf{Description}  & \textbf{Category} & \textbf{\#Vuln. Instance} & \textbf{Impact} &  \textbf{Newly Obtained CVEs/ Acknowledgements}  \\ 
		\hline

        \textbf{B-E1} & Stops advertising if \connectionreq \fontBLEVulTable with Interval/channelMap field set to 0 is sent \cite{blediff,sweyntooth}  & \categoryone \categorytwo & 3 & Crash & Acknowledged by MediaTek  as low severity\\ \hline
        \textbf{B-E2} & Stops advertising with plaintext \encpauseresp \cite{blediff} &  \categorytwo &  2 & Crash &  Acknowledged by MediaTek \& Samsung  as low severity\\ \hline
        \textbf{B-E3} & Bypassing passkey entry in legacy pairing \cite{blediff}  & \categoryone\categorytwo  & 10 & Privacy/ Null Encryption &  Medium severity CVE from Samsung \\ \hline
        % \textbf{B-E4} & Accepts plaintext \encpausereq \fontBLEVulTable before pairing is complete \cite{blediff}  & \categoryone  & 1 &   &   & DoS &  Acknowledged by MediaTek \& Samsung \\ \hline
        \textbf{B-E4} & Accepts malformed \connectionreq \fontBLEVulTable message with increased length \  & \categorythree &  8 & Memory Leakage &  Medium severity CVE from Samsung \\ \hline
        \textbf{B-E5} & Accepts \pairconfirm \fontBLEVulTable with wrong values \cite{blediff} & \categoryone  & 1 & DoS & -  \\ \hline
        \textbf{B-E6} &  \lengthreq \fontBLEVulTable with MaxRxOctets and MaxTxOctets fields set to 1 cause DoS & \categorythree  &  2 & DoS & Acknolwedged by MediaTek as low severity\\ \hline
        \textbf{B-E7} & Unexpected Data Physical Channel PDU during encryption start procedure cause DoS  & \categorythree  &  11 & DoS &  Acknolwedged by Samsung  \\ \hline
        \textbf{B-E8} & \keyexchange \fontBLEVulTable in legacy pairing leading to DoS \cite{sweyntooth} & \categoryone  &  1 & DoS & Reported by Sweyntooth \cite{sweyntooth} and fixed in newer version  \\ \hline 
        \textbf{B-E9} & Two continuous \pairreq \fontBLEVulTable leading to DoS  & \categorythree  &  2 & DoS &  CVE from Microchip with low severity \\ \hline
        \textbf{B-E10} & Plaintext \encpausereq \fontBLEVulTable cause DoS  & \categorythree  &  4 & DoS &  Acknolwedged by MediaTek \& Samsung  as low severity\\ \hline
        \textbf{B-E11} &  \chmreq \fontBLEVulTable with unchanged channelMap field cause DoS  & \categorythree  &  6 & DoS & -  \\ \hline
        
        \textbf{B-I1} & Issue with OOB Pairing Fails \cite{blediff}  & \categorytwo  &  8 & Fingerprinting & -  \\ \hline
        % \textbf{B-I2} & Interoperability issue with reject messages \cite{blediff}  & \categoryone  &  2 &   &   & Fingerprinting &   \\ \hline
        
        % \textbf{B-O1} & Accepts malformed \connectionreq  \fontBLEVulTable message with Timeout field set to 0 \cite{sweyntooth} & \categoryone & 2 &   &   & Fingerprinting &   \\ \hline
        % \textbf{B-O1} & Accept malformed \versionreq \fontBLEVulTable with LLID field set to 0 \cite{sweyntooth} & \categorytwo  &  2 & Fingerprinting & -  \\ \hline
        % \textbf{B-O3} & Accepts malformed \pairreq \fontBLEVulTable message with overflow key size \cite{blediff} & \categorytwo & 1 &   &   & Buffer overflow and possible encryption bypass & Existing from CVE-2019-19196 (6.5 medium severity) by MITRE  \\ \hline
        \textbf{B-O1} & Accept \connectionreq \fontBLEVulTable with Hop field $>$ defined range & \categorythree  &  9 & Fingerprinting & Acknowledged by MediaTek as a functional bug  (Negligible Security Impact)\\ \hline
        
        \textbf{B-O2} & Accept \versionreq \fontBLEVulTable multiple times \cite{sweyntooth} & \categorytwo  &  2 & Fingerprinting & Acknowledged by Samsung \\ \hline
        \textbf{B-O3} & Accept \encreq \fontBLEVulTable with non-zero EDIV and Rand \cite{sweyntooth} & \categorytwo  & 12 & Fingerprinting & Acknowledged by MediaTek as a functional bug  (Negligible Security Impact)\\ \hline 
        
	\end{tabular}
    }
        %\captionsetup{justification=centering}
       
        \vspace*{-0.1cm}
	\label{tab:ble-attacks}
\end{table*}

\noindent\textbf{Attacker model.}
We also assume the Dolev-Yao attacker model for BLE. 
Similar to previous works ~\cite{blediff,blesa,antonioli2020bias},  
the attacker acts as a malicious central and can intercept, replay, modify, or drop packets. The attacker only knows the public information of the target peripheral (e.g., Bluetooth name, address, protocol version number, and capabilities) and does not require knowledge regarding any secret keys shared between the target peripheral and any other device. For issues \textbf{B-E1, B-E2, B-E4, B-I1, B-O1} to \textbf{B-O3}, the attacker can directly establish a connection with the victim peripheral to launch the attacks. For all the other issues, the attacker must inject malicious traffic to the target peripheral ~\cite{cayre2021injectable} when the target pairs with another device for the first time. 
%For all other issues, the attacker must inject malicious traffic into the target peripheral during the target device's initial pairing with another device~\cite{cayre2021injectable}.   

\noindent{\textbf{B-E9: Accepts two continuous \pairreq leading to DoS.}}
The affected peripherals cannot complete the pairing procedure when a malicious central sends two consecutive \pairreq messages. On the other hand, Microchip's BLE device responds to both \pairreq with \pairresp. For other devices not having this vulnerability, the second \pairreq is ignored, and the pairing procedure proceeds as usual.

\noindent\textbf{Impact.} 
This attack leads to DoS, where the victim cannot complete the pairing process. If the central device does not have an auto-reconnect feature, the pairing procedure needs to be manually re-initiated, and even with auto-reconnect, the device needs to restart the entire pairing procedure again. Also, the vulnerability can be exploited to perform a manufacturer-level fingerprinting of vulnerable implementations. Microchip assigned a CVE with low severity to this issue.

\noindent\textbf{\textbf{B-E1}, \textbf{B-E2}, \textbf{B-E6}, \textbf{B-E7}, \textbf{B-E10} and \textbf{B-E11}.}
To exploit these vulnerabilities, i.e., \textbf{B-E6}, \textbf{B-E7}, \textbf{B-E9} to \textbf{B-E11}, the adversary assumptions are identical to those for \textbf{B-E9}. 
For \textbf{B-E6}, we observe that before the pairing process starts, the affected devices will respond to \lengthreq with MaxRxOctets and MaxTxOctects field set to 1, and as a result, the subsequent pairing process cannot be completed. For \textbf{B-E7}, we observe that sending Data Physical Channel PDUs, such as \mtureq, \lengthreq, \versionreq, after \encreq, can cause the peripheral to disconnect the link. For \textbf{B-E10}, we observe that the affected devices cannot finish pairing if it receives a \pauseencreqplaintext after \keyexchange. 
For \textbf{B-E11}, an affected device disconnects after receiving a \chmreq with the unchanged \textit{Channel Map} field. For \textbf{B-E1} and \textbf{B-E2}, the device cannot recover by itself and requires manual reboot. For the other identified DoS attacks, the pairing procedure needs manual re-initiation. Also, sending the packets repeatedly can lead to prolonged DoS and battery depletion for each case.

\noindent{\textbf{B-E4: Accepts malformed \connectionreq with increased data length field value.}}
An implementation accepts \connectionreq with an increased \textit{Data Length} value. The \connectionreq is extended to 247 bytes when \textit{Data Length} field value is increased. Thus, after accepting this packet, the implementation may allocate more memory than required.

\noindent\textbf{Impact.} 
Since the vulnerable device accepts L2CAP packets with the wrong length, more bytes than expected are allocated in memory for an incorrectly implemented device. Sweyntooth ~\cite{sweyntooth} found a similar vulnerability for \pairreq packet, leading to memory leakage. Moreover, an attacker can fingerprint the device at the manufacturer's level, which may be further exploited by abusing other known firmware-level vulnerabilities. Samsung acknowledged this issue and provided a medium-severity CVE.

\noindent{\textbf{B-E3: Bypassing passkey-entry during legacy pairing.}}
During the passkey-entry association method,  
a malicious central can bypass the passkey entry step by sending a \pairrandom message with a temporary key set to 0. This bypasses all MitM protection mechanisms inherent in the passkey entry method. BLEDiff ~\cite{blediff} initially identified this issue in older devices. However, we identified this issue on several newer devices, including peripherals from Samsung, Google, Hisense, and Motorola (see Table \ref{tab:device_info_BLE} in Appendix). 
Samsung assigned a CVE with medium severity to this issue. 

\noindent\textbf{Other Issues.}
\system{} also identified several issues, such as accepting malformed \connectionreq with a Hop field greater than the defined range (\textbf{B-O1}), accepting \versionreq message multiple times (\textbf{B-O2}) and accepting \encreq with non-zero EDIV and Rand field value (\textbf{B-O3}).  The attacker can exploit these to fingerprint the affected device.
%\vspace{-0.cm}
\section{Comparison with Existing Works}
\label{sec:eval_comparison_existing_works}

To address \textbf{RQ2}, we compare \system with existing LTE and BLE testing frameworks. We first provide a qualitative comparison with respect to different metrics and then provide empirical comparisons with respect to the number of vulnerabilities detected, coverage, and cumulative vulnerability detection over time.

%\vspace{-0.1cm}
\subsection{Qualitative Comparison}
\label{sec:eval_comparison_existing_works:qualitative_comparison}

We compare \system with existing works with respect to different metrics as shown in Table~\ref{tab:comparison_existing_works}.
Among these works, DIKEUE ~\cite{dikeue}, DoLTEst ~\cite{doltest}, Contester ~\cite{sherlock}, BaseComp ~\cite{basecomp}, BLEDiff ~\cite{blediff} and BLE Blackbox Fuzzing ~\cite{blackbox-fuzzing} perform stateful testing. 
However, DIKEUE, BaseComp, BLEDiff, and Blackbox Fuzzing do not consider any specific properties corresponding to the protocols while generating queries. 
Although LTEFuzz ~\cite{ltefuzz}, DoLTEst and Contester have property-guided generation, LTEFuzz is not stateful, and DoLTEst and Contester do not generate test cases dynamically. 
Also, SweynTooth ~\cite{sweyntooth} does not consider the guidance of properties while generating test cases. 

Further, DoLTEst, LTEFuzz, and SweynTooth do not perform positive testing, and only DIKEUE, DoLTEst, Contester, BLEDiff, and Blackbox Fuzzing have the capabilities to identify interoperability issues. Finally, none of the works except \system consider efficient scheduling of test traces to maximize property violation detection within a fixed testing budget. 
Only \system simultaneously covers all these aspects of testing. In addition, it is the only dynamic framework that can provide control over the amount of test traces being generated and, hence, can be tuned to satisfy a time budget for testing.

\begin{table}[h]
%\vspace{-0.2cm}
\caption{Comparison with existing testing approaches. 
    % budget based testing
    }
\vspace{-0.4cm}
\resizebox{0.95\columnwidth}{!}{%
    \centering
    \begin{tabular}{@{}lccccccc@{}} 
    \toprule
        \thead[l]{\textbf{Approach}} & 
        \thead{\textbf{Dynamic}} & 
        \thead{\textbf{Stateful} \\ \textbf{Testing}} &
        \thead{\textbf{Property} \\ \textbf{Focused} \\ \textbf{Testing}} &
        \thead{\textbf{Positive} \\ \textbf{Test}  \\ \textbf{Cases}} & 
        \thead{\textbf{Time Budget} \\ \textbf{Wise Tuning}} & 
        \thead{\textbf{Interoperability} \\ \textbf{Issue} \\ \textbf{Detection}} \\ 
    \midrule
        DIKEUE \cite{dikeue} & \cmark & \cmark & \xmark & \cmark  & \xmark & \cmark \\

        LTEFuzz \cite{ltefuzz} & \cmark & \xmark & \cmark & \xmark & \xmark & \xmark \\

        DoLTEst \cite{doltest} & \xmark & \cmark & \cmark & \xmark & \xmark & \cmark \\
        
        Contester \cite{sherlock} & \xmark & \cmark & \cmark & \cmark & \xmark & \cmark \\
        
        Basecomp \cite{basecomp} & \xmark & \cmark & \xmark & \cmark  & \xmark & \xmark \\

        BLEDiff \cite{blediff} & \cmark & \cmark & \xmark & \cmark & \xmark & \cmark \\

        SweynTooth \cite{sweyntooth} & \cmark & \xmark & \xmark & \xmark & \xmark & \xmark \\

        Blackbox Fuzzing \cite{blackbox-fuzzing} & \cmark & \cmark & \xmark & \cmark & \xmark & \cmark \\

        \rowcolor{lightgray} \textbf{\system} & \textbf{\cmark} & \cmark & \cmark & \cmark & \cmark & \cmark \\

    \bottomrule
                
    \end{tabular}
    }
    \vspace{-0.1cm}
    \label{tab:comparison_existing_works}
\end{table}

%\vspace{-0.3cm}

\subsection{Total Number of Vulnerability Detection}

\label{sec:eval_comparison_existing_works:total_vulnerability}

{
We compare \system with DIKEUE ~\cite{dikeue}, and DoLTEst ~\cite{doltest} for LTE, and BLEDiff ~\cite{blediff} and Sweyntooth ~\cite{sweyntooth} for BLE, with respect to the total number of identified vulnerabilities without considering any time budget.  
%\textcolor{red}{We enumerate all the identified issues for each existing work and also all the identified issues of \system.} 
For each vulnerability, we determine whether the vulnerability is identifiable by-- (i) only \system, (ii) only the baseline work, or (iii) both works. For DIKEUE ~\cite{dikeue} and BLEDiff ~\cite{blediff}, we used their provided alphabet.
% for testing. 

}

\begin{table}[t]
\caption{Detected vulnerability count by existing works.}
\vspace{-0.3cm}
\resizebox{\columnwidth}{!}{
\centering
\begin{tabular}{|c|
>\centering m{3.9cm}|
c|
c|}
\hline
% Ishtiaq: made the header 3 lines because 2 lines make them too small and unreadable.
\textbf{Baseline} & \textbf{\begin{tabular}[c]{@{}c@{}}\#Vuln. both \system \\ and Baseline can Identify\end{tabular}} & \textbf{\begin{tabular}[c]{@{}c@{}}\#Vuln. only \system \\ can Identify\end{tabular}} & \textbf{\begin{tabular}[c]{@{}c@{}}\#Vuln. only Baseline \\ can Identify\end{tabular}} \\ \hline
DIKEUE            & 15                                                                                      & 4                                                                                        & 0                                                                                         \\ \hline
DoLTEst           & 25                                                                                      & 5                                                                                        & 1                                                                                         \\ \hline
BLEDiff           & 13                                                                                      & 7                                                                                        & 0                                                                                         \\ \hline
Sweyntooth         & 18                                                                                      & 8                                                                                        & 2                                                                                         \\ \hline
\end{tabular}
}
\label{tab:vul_count} 
% \tianwei{the number of ble part of this table will be updated after we ensure potential new vulnerabilities}
\vspace{-0.4cm}
\end{table}

{
We present the results in Table \ref{tab:vul_count}. Our analysis reveals that \system can identify all vulnerabilities detected by the existing works except for one identified by DoLTEst ~\cite{doltest}. To detect this particular vulnerability, DoLTEst manually crafts and sends two traces to \pimp: one with an \idreq message using security header 12 and another using security header 15, observing any differences in the target's response. In contrast, \system can automatically generate and reason about only one trace at a time, meaning it can only check if a single trace violates any security properties at once. \system is not designed to reason about hyper-properties which require simultaneous consideration of multiple traces.
}

{In contrast, even without considering a testing budget, DoLTEst \cite{doltest} cannot detect replay vulnerabilities (e.g., \textbf{L-E3, L-E4}) and also stateful vulnerabilities where the vulnerability is triggered at a state not defined by it. For example, issue \textbf{L-E8} can only be triggered after authentication and before NAS security activation, but DoLTEst does not incorporate this state. Sweyntooth ~\cite{sweyntooth} cannot detect logical vulnerabilities because it lacks a test oracle to identify such issues. Furthermore, to detect all vulnerabilities identified by \system through an automata-based learning approach ~\cite{dikeue,blediff}, we must incorporate a set of alphabet containing all symbols used by Proteus, i.e., all messages, message fields, and their mutations during PSM learning. Using this alphabet in a 4G LTE environment (111 symbols) with DIKEUE ~\cite{dikeue} would require 87,246 unique queries ($\sim$133 days if tested OTA) to learn the PSM of an implementation, taking over four months to detect all vulnerabilities \system identified with 3000 queries (2 days when tested OTA). 
}

{
Additionally, as shown in Table \ref{tab:vul_count}, \system detects 4-5 new vulnerabilities in LTE and 7-8 new vulnerabilities in BLE compared to these prior works. Thus, \system is more effective in identifying vulnerabilities in the tested protocols.
}

\subsection{Vulnerability Count Growth}
\label{sec:eval_comparison_existing_works:cvul_over_time}
We demonstrate \system{} 's efficiency in detecting vulnerabilities over time. This evaluation also demonstrates the quality of the generated traces by \system to detect vulnerabilities.
We compare \system with the existing works in terms of cumulative vulnerability count on 3 devices from BLE and 2 devices on LTE.
Further, since we focus on logical bugs, we consider the security properties as an oracle to detect vulnerabilities. Note that we did not observe any false positives (i.e., any reported property violation that does not result in exploitable vulnerability). The default value of the length budget parameter for \dpalgo is $l+1$, where $l$ is the number of observations in any test skeleton for properties. Also, we set the default mutation budget to 2.

\noindent\textbf{Cumulative vulnerability count on BLE devices.} 
We compute the cumulative vulnerability count obtained over time on three BLE devices-- Galaxy S6, Galaxy A22, and Oppo Reno7,  with both \system and BLEDiff ~\cite{blediff}. 
We consider BLEDiff as our baseline. However, BLEDiff learns the PSM of the target device first and then performs differential testing. 
Thus, to make a fair comparison, we consider the queries generated during learning as input queries to detect vulnerabilities and use the security properties obtained for \system as an oracle to determine whether any vulnerability is detected. 
We run \system and BLEDiff for 24 hours and count the number of detected vulnerabilities. {We run the experiments for 3 iterations and present their average in Figure \ref{fig:cvul_ble_blediff}.}

\begin{figure}[htbp]
\begin{minipage}[t][][b]{0.32\linewidth}
    \centering
    \includegraphics[width=\linewidth]{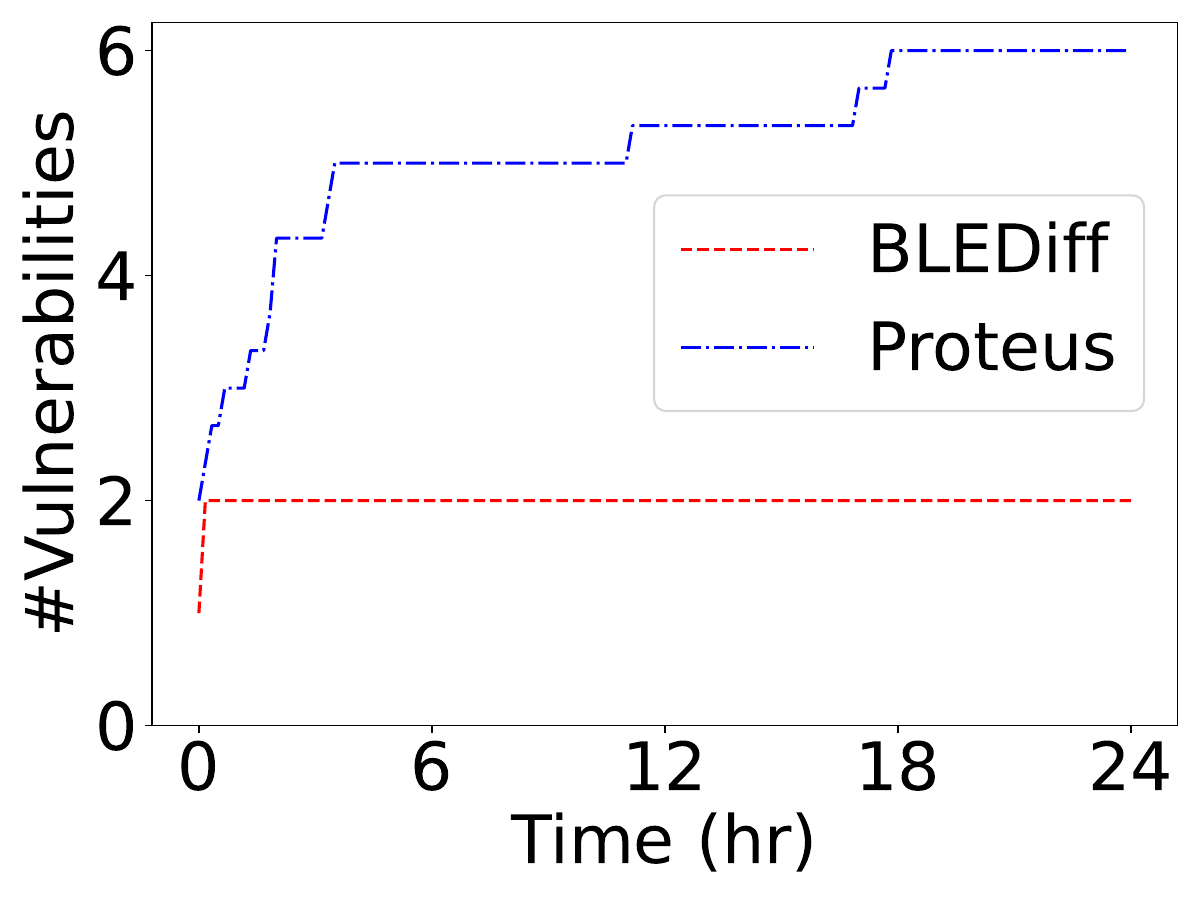}
    \vspace{-0.75cm}
    \caption*{(a) Galaxy A22}
    % \label{fig:}
\end{minipage}
% \hspace{1mm}
\begin{minipage}[t][][b]{0.32\linewidth}
    \centering
    \includegraphics[width=\linewidth]{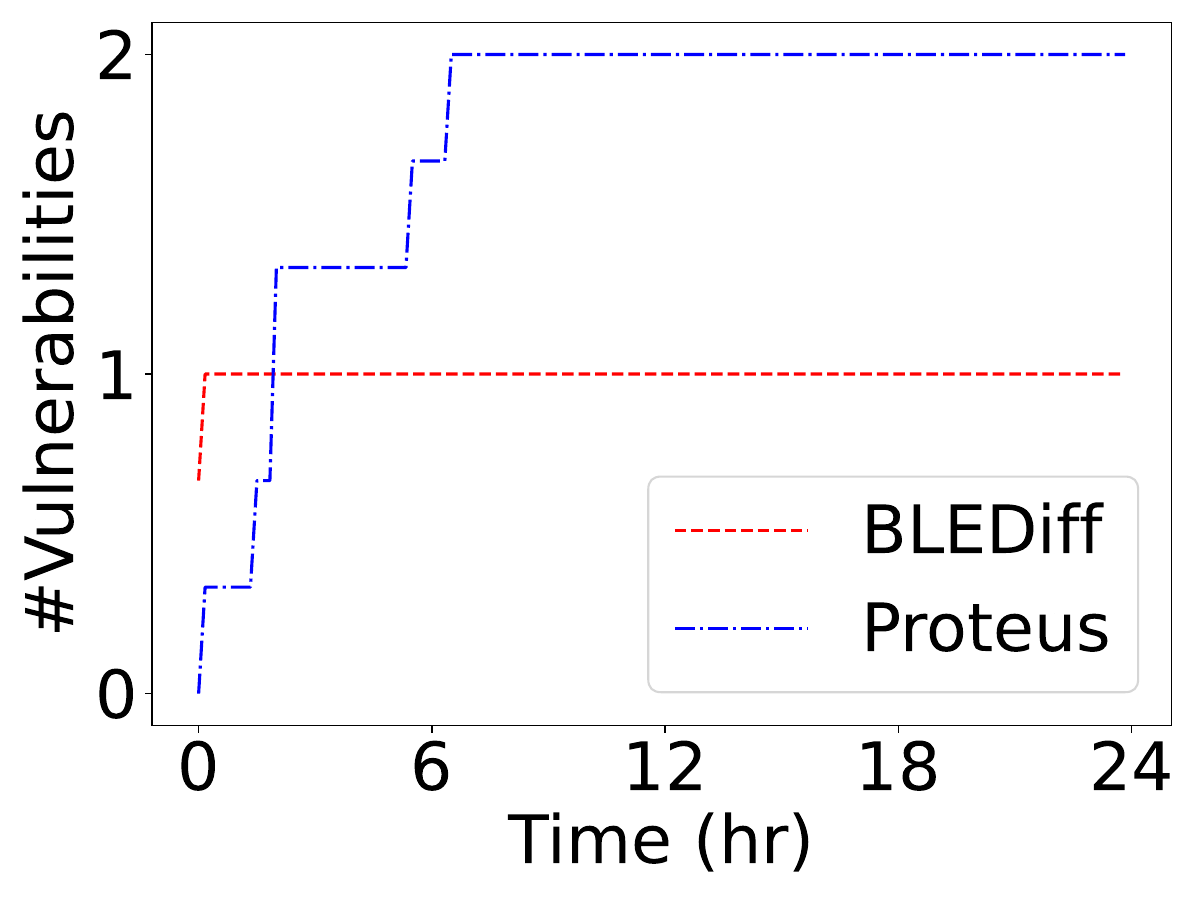}
    \vspace{-0.75cm}
    \caption*{(b) Galaxy S6}
    % \label{fig:}
\end{minipage}
\begin{minipage}[t][][b]{0.32\linewidth}
    \centering
    \includegraphics[width=\linewidth]{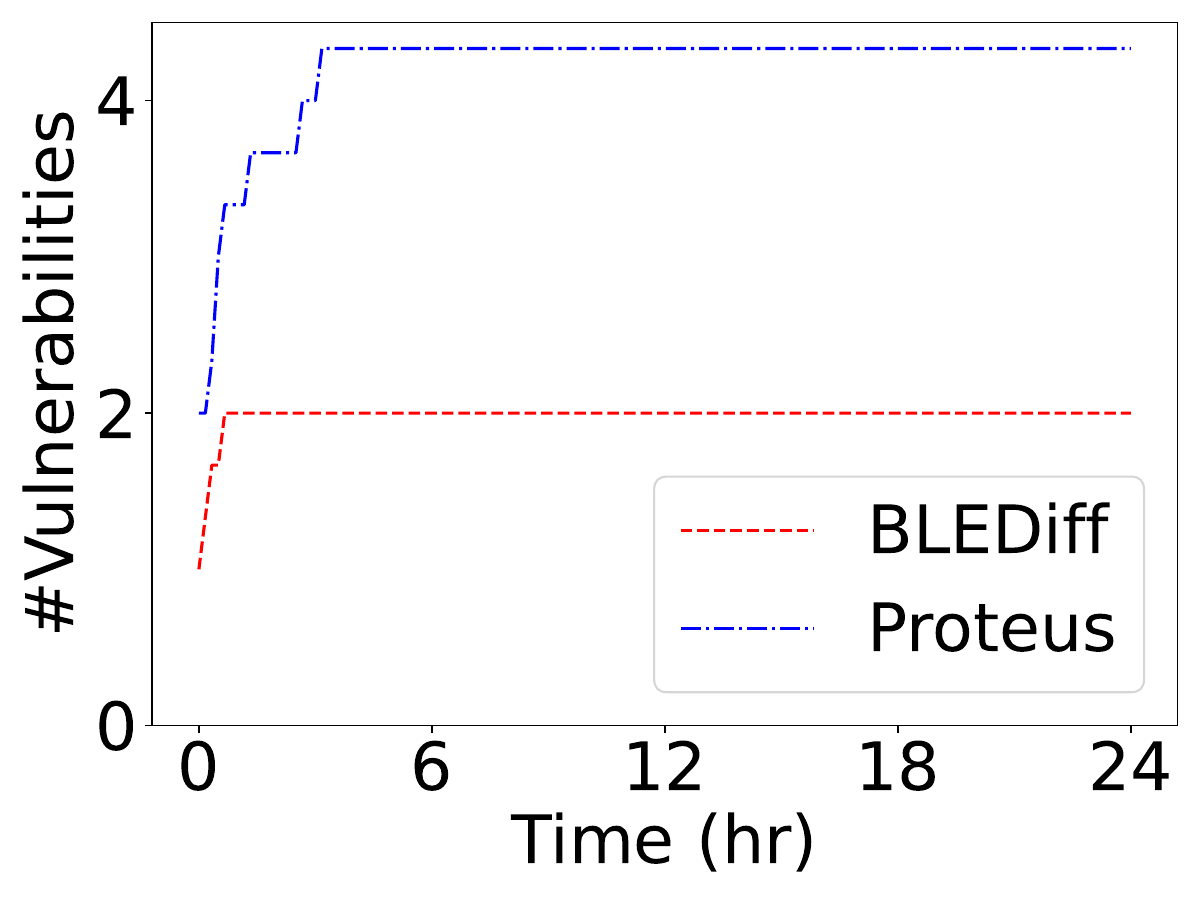}
    \vspace{-0.75cm}
    \caption*{(c) Oppo Reno7}
    % \label{fig:}
\end{minipage}
% \hspace{1mm}
   % \vspace{-0.3cm}
    \caption{Cumulative vulnerability comparison in BLE.}
    \label{fig:cvul_ble_blediff}
    %\vspace{-0.4cm}
\end{figure}

Figure \ref{fig:cvul_ble_blediff} shows that \system can detect more than 5 vulnerabilities on average within 24 hours, whereas the learning queries from BLEDiff can only detect a maximum of 2 vulnerabilities in the tested devices. 
Although \system, at first, falls behind BLEDiff in speed in S6, 
upon inspection, we find that the only vulnerability identified by BLEDiff is a trivial one, requiring a single message, and \system later finds more vulnerabilities, which BLEDiff cannot detect in 24 hours. 
In other cases, \system consistently performs better than BLEDiff. 
This demonstrates that \system is more efficient in detecting vulnerabilities than BLEDiff's automata learning-based approach, suggesting a better quality of the generated traces. 

\noindent\textbf{Cumulative vulnerability count on LTE devices.} For LTE, we consider Pixel7 and Huawei P8 Lite and compute the cumulative vulnerabilities detected over time by both Proteus and the baseline works-- DoLTEst ~\cite{doltest}, DIKEUE ~\cite{dikeue}. 
We run our experiment for $\sim$ 7 hours since all DoLTEst queries are executed within that period. 
Again, we run each test for 3 iterations and report the average count in Figure \ref{fig:cvul_lte_doltest}.

\begin{figure}[htbp]
\begin{minipage}[t][][b]{0.4\linewidth}
    \centering
    \includegraphics[width=\linewidth]{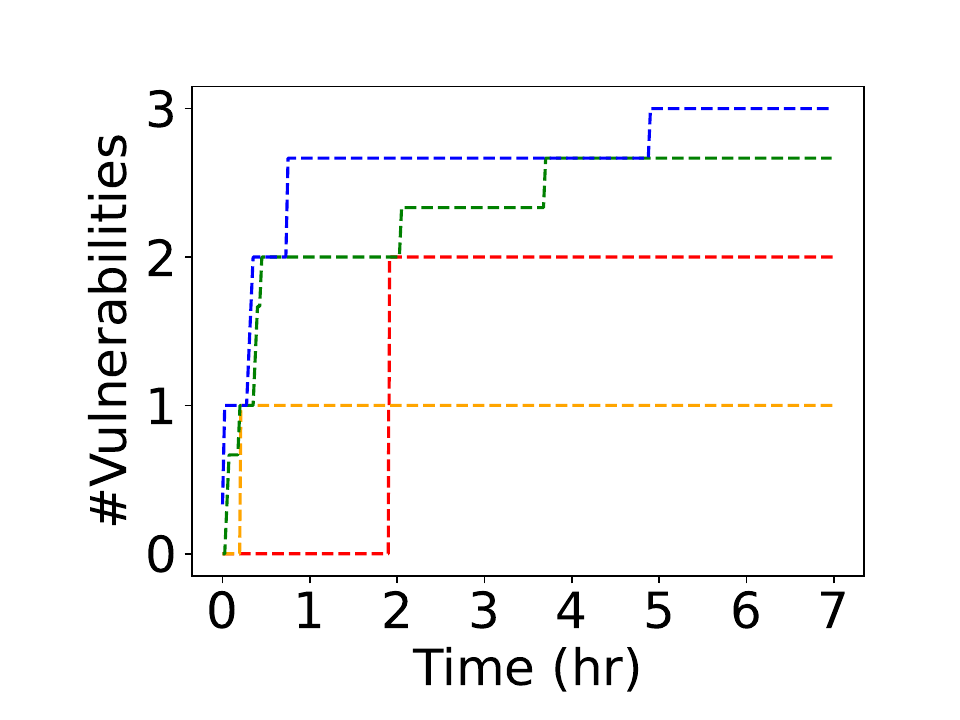}
     \vspace{-0.7cm}
    \caption*{(a) Pixel 7}
    % \label{fig:}
\end{minipage}
\begin{minipage}[t][][b]{0.4\linewidth}
    \centering
%    \vspace{-0.7cm}
    \includegraphics[width=\linewidth]{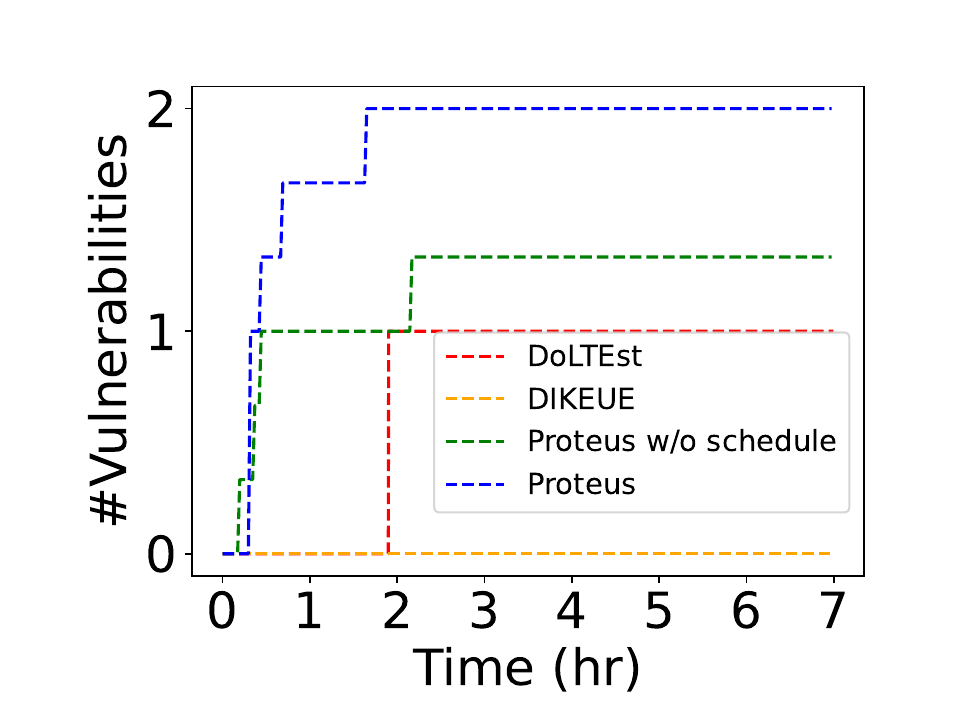}
    \vspace{-0.7cm}
    \caption*{(b) Huawei P8 Lite}
    % \label{fig:}
\end{minipage}
% \hspace{1mm}
    %\vspace{-0.3cm}
    \caption{Cumulative vulnerability comparison in LTE.}
    \label{fig:cvul_lte_doltest}
    % \vspace{-3mm}
\end{figure}

Figure \ref{fig:cvul_lte_doltest} shows that 
\system detects 3 and 2 vulnerabilities on average within 7 hours in Pixel7 and Huawei P8lite, respectively, whereas DoLTEst detects 2 and 1 on average. 
Moreover, the figure presents that \system detects those vulnerabilities faster than DoLTEst and DIKEUE, representing its superior efficiency.  

\subsection{Coverage Growth}
This experiment aims to evaluate the efficacy of the generated traces with respect to coverage growth. 

\begin{figure}[htbp]
\begin{minipage}[t][][b]{0.4\linewidth}
    \centering
    \includegraphics[width=\linewidth]{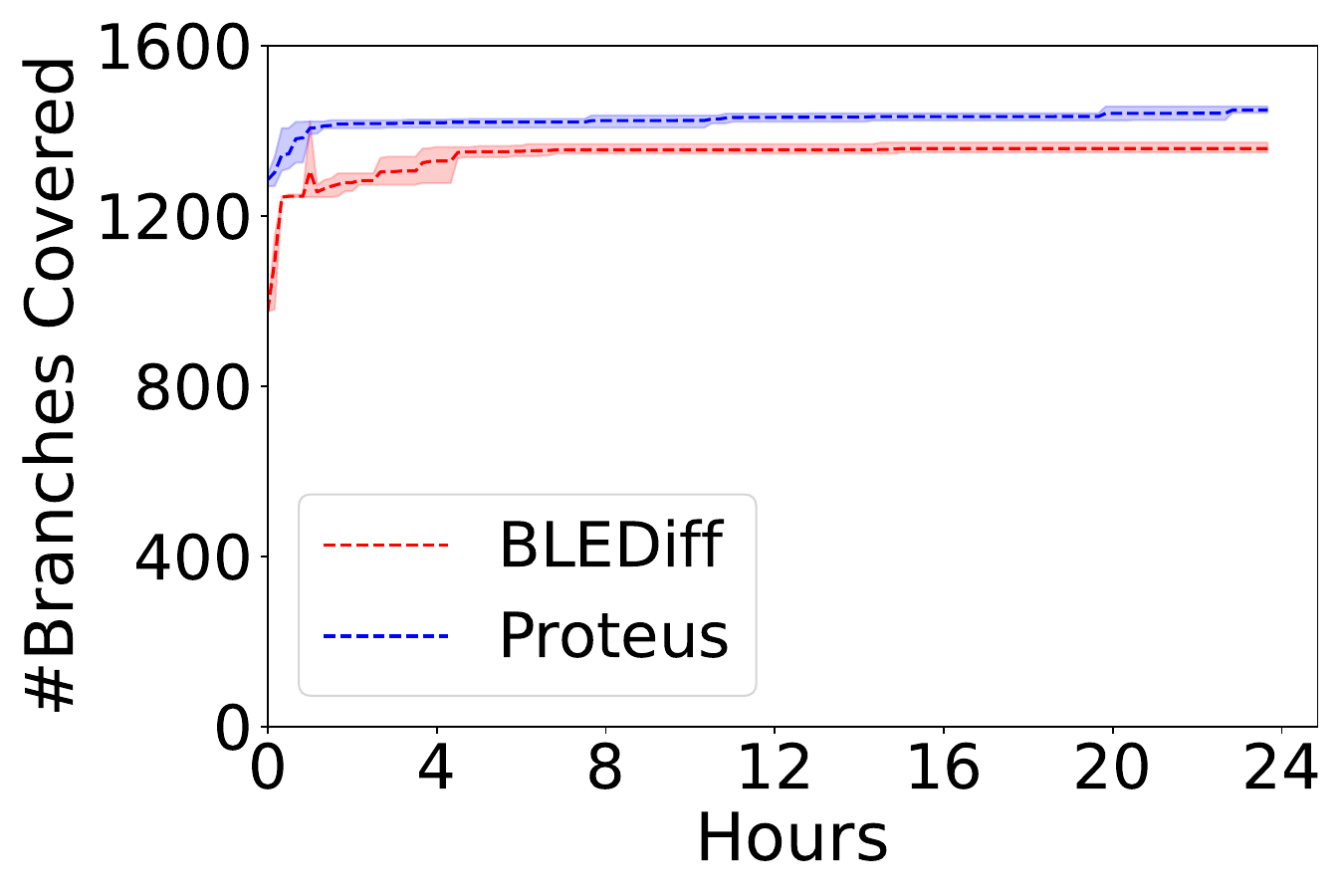}
    \vspace{-0.75cm}
    \caption*{(a) BLE coverage}
    % \label{fig:}
\end{minipage}
% \hspace{1mm}
\begin{minipage}[t][][b]{0.4\linewidth}
    \centering
    \includegraphics[width=\linewidth]{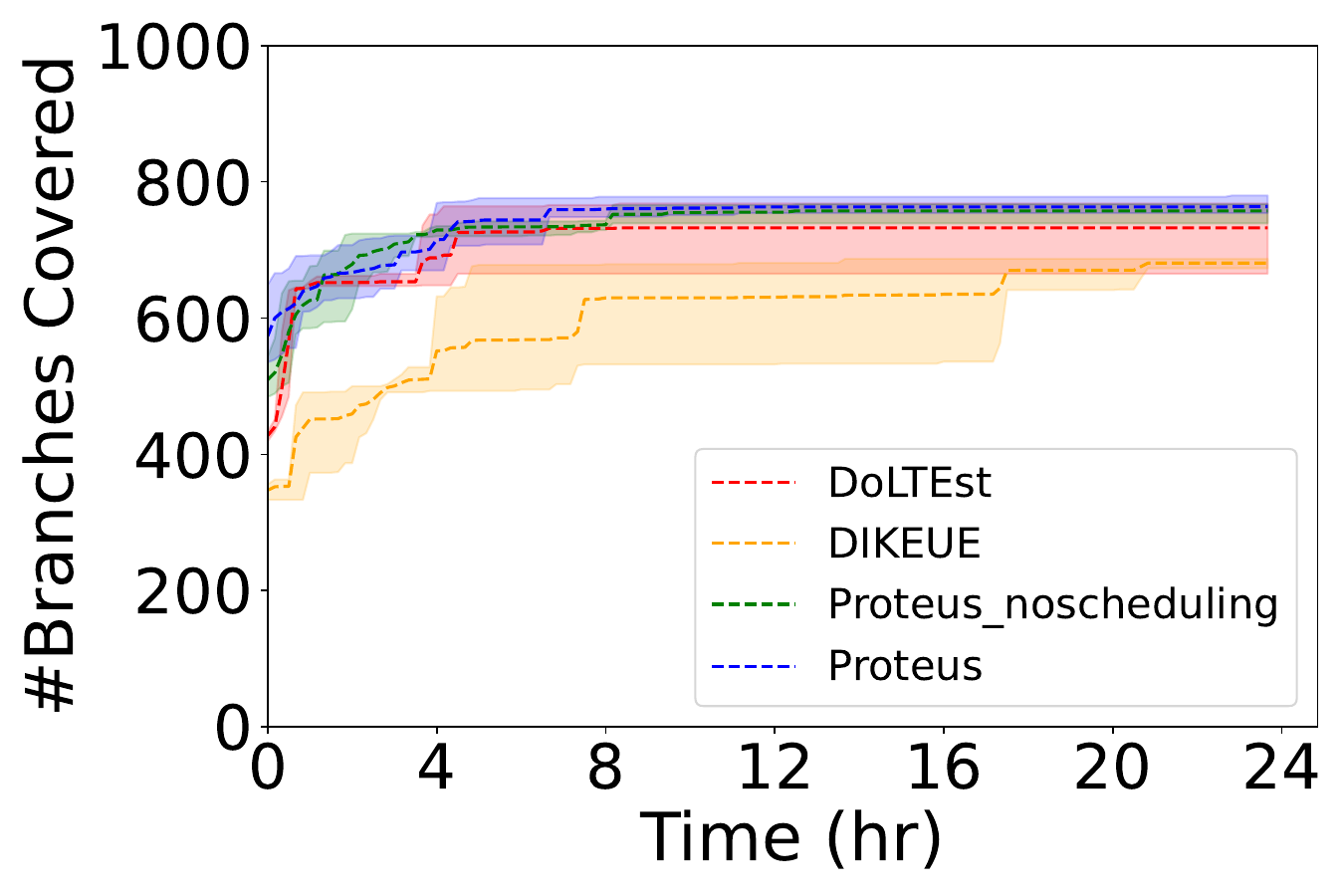}
    \vspace{-0.75cm}
    \caption*{(b) LTE coverage}
    % \label{fig:}
\end{minipage}
    %\vspace{-3.5mm}
    \caption{Coverage growth over time.}
    \label{fig:both-cov}
     \vspace{-2mm}
\end{figure}

\noindent\textbf{Coverage growth in BLE devices.}
At first, we compare \system against BLEDiff ~\cite{blediff} with respect to the number of branches covered over time while testing BLE implementations. 
Similar to \S\ref{sec:eval_comparison_existing_works:cvul_over_time}, we consider the queries used for learning PSM of the target implementation as inputs to BLEDiff. 
{
Using BTStack v1.5.6.3 ~\cite{btstack} as the target, we perform three 24-hour runs of both \system and BLEDiff to determine average branch coverage.

}
We present the results in Figure \ref{fig:both-cov} (a), which
shows that \system achieves $\sim10\%$ more coverage 3 times faster than BLEDiff.

\noindent\textbf{Coverage growth in LTE devices.}
For LTE, we compare \system against DIKEUE ~\cite{dikeue} and DoLTEst ~\cite{doltest}. {We run each tool 3 times on srsUE ~\cite{srsRAN}, with each run lasting 24 hours, and report their average branch coverage over time.}
Figure \ref{fig:both-cov}(b) shows the coverage growth in LTE devices, where it is evident that \system achieves more branch coverage faster than both the baselines.
% \vspace{-0.2cm}

\section{Performance of \system}
\label{sec:eval_performance_proteus}
\begin{table}[ht]
\caption{Regular expression representing acceptance of a replayed \gutiral and \smcmd}
\vspace{-0.2cm}
\centering
\resizebox{1\linewidth}{!}{
\begin{tabular}{|c|p{10cm}|}
\hline
\textbf{Property} & \textbf{Expression} \\
\hline
$\sigma_g$ &  (\enablesone/ \attachreq) (\authreq/ \authresp) $(.)^{\ast}$ (\smcmd / {\footnotesize\textsc{security\_mode \_complete}\normalsize\xspace}) $(.)^{\ast}$ (\rrcsmcmd / {\footnotesize\textsc{rrc \_ security \_mode \_complete}\normalsize\xspace}) $(.)^{\ast}$ (\attachacc/ \attachcomp)  $(.)^{\ast}$
(\gutiral / {\footnotesize\textsc{guti \_ reallocation \_ complete}\normalsize\xspace}) $(.)^{\ast}$ (\gutiralrep / {\footnotesize\textsc{guti \_ reallocation \_ complete}\normalsize\xspace})   \\
\hline
$\sigma_s$ & (\enablesone/ \attachreq) (\authreq/ \authresp) $(.)^{\ast}$ (\smcmd / {\footnotesize\textsc{security\_mode \_complete}\normalsize\xspace}) $(.)^{\ast}$ (\smcmdreplay / {\footnotesize\textsc{security\_mode \_complete}\normalsize\xspace})
 \\
\hline
\end{tabular}
}
\label{tab:security_properties}
\end{table}

\noindent\textbf{Effectiveness of considering both security property and guiding PSM.} We compare \system with two approaches described in \S\ref{sec:motivation}: (i) \textit{PSM only approach}, where we sample traces from our guiding PSM and perform mutations without any guidance (ii) \textit{Property-only approach,} where we consider a RE representing the violation of the security property, and instantiate the RE using arbitrary observations independent of any guidance from a PSM. We use an RE representing the acceptance of a replayed \gutiral attack after a successful registration procedure provided in Table \ref{tab:security_properties}.
We test a HiSense F50+ device to check the number of queries it takes to detect the vulnerability using the two approaches and \system{}. We observe that \system and the property-only approach take 366 and 1690 queries to detect the vulnerability, respectively. The PSM-only approach cannot detect the vulnerability in 3000 queries. This signifies that combining the guidance from both PSM and a security property assists in quickly discovering the vulnerability.

\noindent\textbf{Effect of varying mutation and length budget.} We examine how the number of generated instantiated traces varies according to the length and mutation budget we set in \dpalgo. We consider the two regular expressions $\sigma_g$ and $\sigma_s$ (Table \ref{tab:security_properties}), representing the acceptance of replayed \gutiral and \smcmd, respectively, as text skeletons.

To observe the effect of varying the length budget, we set the mutation budget to 2 and varied the length budget from 4 to 10. The number of unique instantiated traces generated for both the test skeletons are presented in Table \ref{tab:eval_length_budget}. We observe that the length budget has to have a minimum length equal to at least the number of literals present in the trace skeleton since, for a lower length budget, no sequence would be able to satisfy the given skeleton. On the other hand, the total number of unique traces generated increases as the length budget increases.

To observe the effect of varying the mutation budget, we vary it from 1 to 3 for both trace skeletons. We observe that if the length budget is constrained, e.g., a length budget of 5 with 4 literals in $\sigma_s$, then even with a higher mutation budget, we would not have many more test cases since there would be no space to inject any mutation. However, if the length budget is increased, the number of traces generated increases with the increase of the mutation budget. However, we empirically observed that a mutation budget of more than 2 did not yield any extra vulnerabilities. 

\begin{table}[t]
\caption{No. of traces generated for varying mutation budget.
}
\vspace{-0.3cm}
\centering
\resizebox{0.85\columnwidth}{!}{
\begin{tabular}{|ccc||c||ccc|}
\hline
\multicolumn{1}{|c|}{\textbf{Property}} & \multicolumn{1}{c|}{\textbf{\begin{tabular}[c]{@{}c@{}}Number of \\ Observations\end{tabular}}} & \textbf{\begin{tabular}[c]{@{}c@{}}Number of\\ Wildcards\end{tabular}} & \textbf{\begin{tabular}[c]{@{}c@{}}Length\\ Budget\end{tabular}} & \multicolumn{3}{c|}{\textbf{Mutation Budget}}                                  \\ \hline
\multicolumn{3}{|c||}{}                                                                                                                                                                                           &                                                                  & \multicolumn{1}{c|}{\textbf{1}} & \multicolumn{1}{c|}{\textbf{2}} & \textbf{3} \\ \hline
\multicolumn{1}{|c|}{$\sigma_s$}             & \multicolumn{1}{c|}{4}                                                                      & 2                                                                       & 5                                                                & \multicolumn{1}{c|}{22}         & \multicolumn{1}{c|}{53}         & 63         \\ \hline
\multicolumn{1}{|c|}{$\sigma_g$}           & \multicolumn{1}{c|}{7}                                                                      & 5                                                                        & 8                                                                & \multicolumn{1}{c|}{49}        & \multicolumn{1}{c|}{121}        & 166        \\ \hline
\multicolumn{1}{|c|}{$\sigma_g$}           & \multicolumn{1}{c|}{7}                                                                      & 5                                                                        & 9                                                               & \multicolumn{1}{c|}{1441}      & \multicolumn{1}{c|}{5662}      & 11560      \\ \hline
\end{tabular}
}

\label{tab:eval_mutation_budget}
%\vspace{-0.5cm}
\end{table}

\noindent\textbf{Efficiency of scheduling approach.} We compare with a version of \system that randomly schedules properties and instantiated traces (denoted as \system w/o scheduling). The experimental setup is the same as \system. Figure \ref{fig:both-cov} and \ref{fig:cvul_lte_doltest} show its coverage and cumulative vulnerability count over time. We observe that with our scheduling approach, we reach a higher coverage and detect vulnerabilities faster.
\section{Discussions}\label{sec:discussions}

\noindent\textbf{Scope of \system.} 
In addition to logical bugs, \system can 
conceptually cause crashes of the COTS device under test  
due to memory bugs (\eg, use-after-free). 
As \system operates in a black box setting, it  
cannot transparently observe inside the analyzed devices to discern crash bugs. 
As such, we consider such crash-inducing bugs to be outside \system{}'s  
scope.

\noindent\textbf{Proteus's reliance on state machine and desired properties.} 
As discussed before, \system relies on having access to the 
protocol's state machine and its desired security and privacy properties 
to generate meaningful test cases. One can consider this a limitation  
since one has to obtain both the state machine and properties of a protocol 
before testing of its implementation can commence. In our evaluation, we, however, demonstrate 
that access to this extra information is well worth the effort as it enables \system to 
generate high-quality test cases within a given testing budget.

Prior works applied formal 
verification techniques to evaluate the security and privacy of different 
protocol designs ~\cite{lteinspector, 5greasoner}, where both protocol models and properties are manually constructed. 
For protocols where such efforts are underway, \system can take advantage of the constructed models and properties. Another approach currently gaining traction is using natural language processing (NLP) techniques and large language models to automatically extract protocol state machines from the standard specification text ~\cite{rfcnlp, hermes}. 
To evaluate the feasibility of applying such approaches, 
we used one such tool, Hermes ~\cite{hermes}, to automatically extract the protocol state machine of LTE from the standard specification. After comparing the extracted state machine with the hand-constructed one we use for our evaluation, we observe that the extracted state machine is missing a transition. Conceptually, \system can operate with a state machine that is not entirely accurate at the cost of some spurious test cases. 
As a thought exercise, we decided to use \system with the HERMES-extracted 
state machine to test using \system with the same security properties and experimental setup. We ran \system against Pixel 7 for 24 hours and identified all 3 detected issues identified by \system using an entirely correct PSM. 
Despite the missing transition in the automatically generated state machine, 
we observed that \system was able to identify all the vulnerabilities that 
it uncovered when using the hand-constructed state machine. This 
corroborates our hypothesis of \system{}'s loose reliance on the accuracy 
of the state machine.

\noindent\textbf{Limitations of open-source adapter implementations impose constraints on testing.} \system's capabilities are constrained by the message types and predicates that we can implement in the \textit{adapter} (\S\ref{sec:trace_dispatcher}) with open-source protocol stacks, which may support only a limited set of message types and predicates. Moreover, as a black-box testing method, \system can only observe the output messages of the IUT. Therefore, we only include transitions in our guiding PSM, for which the implementation should provide observable responses according to the specifications.
%\vspace{-1mm}
\begin{table}[t]
\caption{Number of unique traces generated with various length budget values.}
\vspace{-0.3cm}
\resizebox{\columnwidth}{!}{
\centering
% \scriptsize
\fontsize{8}{8}\selectfont
\begin{tabular}{|ccc||ccccccc|}
\hline
\multicolumn{1}{|c|}{\textbf{Property}} & \multicolumn{1}{c|}{\textbf{\begin{tabular}[c]{@{}c@{}}\# Literals\end{tabular}}} & \textbf{\begin{tabular}[c]{@{}c@{}}\# Kleene Star\end{tabular}} & \multicolumn{7}{c|}{\textbf{Length Budget (No. Of Unique Traces)}}                                                                                                                                                                                    \\ \hline
\multicolumn{3}{|c||}{}                                                                                                                                                                & \multicolumn{1}{c|}{\textbf{4}} & \multicolumn{1}{c|}{\textbf{5}} & \multicolumn{1}{c|}{\textbf{6}} & \multicolumn{1}{c|}{\textbf{7}} & \multicolumn{1}{c|}{\textbf{8}}          & \multicolumn{1}{c|}{\textbf{9}}          & \textbf{10}                  \\ \hline

\multicolumn{1}{|c|}{$\sigma_s$}        & \multicolumn{1}{c|}{4}                                                             & 2                                                               & \multicolumn{1}{c|}{1} & \multicolumn{1}{c|}{62} & \multicolumn{1}{c|}{2638} & \multicolumn{1}{c|}{\textgreater{}20000} & \multicolumn{1}{c|}{\textgreater{}20000} & \multicolumn{1}{c|}{\textgreater{}20000} & \textgreater{}20000 \\ \hline

\multicolumn{1}{|c|}{$\sigma_g$}        & \multicolumn{1}{c|}{7}                                                             & 5                                                               & \multicolumn{1}{c|}{0} & \multicolumn{1}{c|}{0}  & \multicolumn{1}{c|}{0}   & \multicolumn{1}{c|}{1}     & \multicolumn{1}{c|}{121}                  & \multicolumn{1}{c|}{5662}                 & \textgreater{}20000              \\ \hline
\end{tabular}
}
% \vspace{-0.4cm}
\label{tab:eval_length_budget}
\end{table}
\section{Related Works}
\noindent\textbf{Protocol testing.} 
Prior works on testing protocol implementations include mutation-based fuzzing approaches ~\cite{pham2020aflnet, fuzzowski, schumilo2022nyx}. However, they require significant time to reach diverse code paths and identify vulnerabilities. 
DY Fuzzing ~\cite{dy_fuzzing} mutates network packets considering a Dolev-Yao adversary in the communication channel and combines domain knowledge for effective protocol testing. 
%As these approaches rely on mutations of packets, 
%compared to \system, these works are inefficient in finding vulnerabilities.
However, since the mutation scheme of these works does not consider the security property being tested or the current protocol state, they are inefficient in finding vulnerabilities compared to \system{}. %\syed{Why?}
Fiterau-Brostean et al. ~\cite{fiterau2023automata} developed a method to first learn an automaton from protocol implementations and then compare it against a catalog of manually obtained bug patterns, represented as deterministic finite automata (DFA) to identify vulnerabilities. However, this approach requires learning the state machine from the implementation, which is costly (\S\ref{sec:experiments}). It is also limited to identifying only known vulnerabilities. 
%New bugs were found primarily because of the application of automata learning to targets not previously tested by automated techniques.
The new bugs found in previously untested targets are mostly instances of known bugs found in other devices earlier.
%\textcolor{red}{Additionally, it is limited to identifying only known vulnerabilities, with new bugs discovered primarily due to applying automata learning on targets that had not been previously tested using automated techniques.}
%
Further, another direction of work employs differential testing to find exploitable and interoperability issues ~\cite{walz2017exploiting, fingerprinting-ble}. 
Because of fixed sets of packets, they cannot explore vulnerabilities that require diverse packet fields and corresponding values.

\noindent\textbf{Security testing of cellular protocols. }
Previous studies investigating the security of LTE implementations have typically focused on either OTA testing ~\cite{doltest, dikeue, ltefuzz, park2016white, Shaik2016} or the analysis of baseband firmware ~\cite{basecomp, basespec, basesafe, baseband-1, baseband-2}. 
Park et al. ~\cite{doltest} have developed a negative testing framework along with a set of extensive test cases to detect vulnerabilities in UE devices. However, these test cases are manually designed and statically generated instead of being generated dynamically depending on the implementations. 
%Due to the limited number of symbols in the alphabet, DIKEUE ~\cite{dikeue} may fail to detect some vulnerabilities. 
Kim et al. ~\cite{ltefuzz} introduce a semi-automated approach for fuzzing the LTE control plane, which relies on some basic security properties. 
Researchers have also used NLP to generate test cases from cellular specifications ~\cite{sherlock, bookworm}.
% Yi Chen et al. ~\cite{sherlock} propose Contester, which innovatively uses natural language processing techniques to automate test case generation and uncovered many implementations that do not adhere to the specification. 

\noindent\textbf{Security testing of BLE protocols.}
Frankenstein~\cite{frankenstein} utilizes advanced firmware emulation techniques to fuzz firmware dumps, enabling the direct application of fuzzed input to a virtual modem. FirmXRay ~\cite{firmxray} proposes static binary analysis tools to find the security issues caused without running the firmware. On the other hand, ToothPicker ~\cite{toothpicker} focuses on one platform, providing host fuzzing techniques for this platform. However, these works do not perform stateful fuzzing. BLESA~\cite{blesa} utilizes ProVerif ~\cite{proverif} to perform formal verification of BLE protocols. InternalBlue ~\cite{internalblue} performs reverse engineering on multiple Bluetooth chipsets to test and find vulnerabilities. BLEScope ~\cite{blescope} identifies misconfigured devices by analyzing the companion mobile apps. SweynTooth ~\cite{doltest} offers a systematic testing framework to fuzz BLE implementations. However, none of these works consider stateful and property-focused testing. BLEDiff ~\cite{blediff} develops an automated black-box protocol noncompliance testing framework that can uncover noncompliant behaviors in BLE implementations. However, their approach cannot control the number of test cases generated and be carried out within a time budget $t$.
%\vspace{-1mm}
\section{Conclusion and Future Work}
%\mukit{TODO: this is the only conclusion from ieee submission.}
We design \system, a black box, protocol state machine and property-guided, and budget-aware automated testing approach for discovering 
logical vulnerabilities in wireless protocol implementations.  % \system which generates etests to detect logical protocol vulnerabilities. 
\system leverages guidance from a PSM and also security properties to efficiently generate traces that can detect more vulnerabilities using much less amount of OTA queries than existing works. %\system leverages a dynamic programming algorithm to generate the traces. 
In the future, we will use \system to analyze other wireless protocols. 

%In the future, we would explore how to integrate our framework with an automatically generated FSM, so that it could be used to test open-source protocol implementations and detect logical vulnerabilities.  
\section*{Acknowledgements}
We thank the anonymous reviewers and the shepherd for their feedback and suggestions. We also thank the vendors for cooperating with us during the responsible disclosure. 
This work has been supported by the NSF under grants 2145631, 2215017, and 2226447, the Defense Advanced Research Projects Agency (DARPA) under contract number D22AP00148, State University of New York’s Empire Innovation Program, and the NSF and Office of the Under Secretary of Defense-- Research and Engineering, ITE 2326898, as part of the NSF Convergence Accelerator Track G: Securely Operating Through 5G Infrastructure Program.

% \begin{acks}
% \end{acks}

\bibliographystyle{ACM-Reference-Format}
\bibliography{proteus}

\clearpage
% %%
% %% If your work has an appendix, this is the place to put it.
\appendix
\section{Brief Description of 4G LTE and BLE Protocols}
\begin{figure}[ht]
    \centering
    \includegraphics[width=0.8\linewidth]{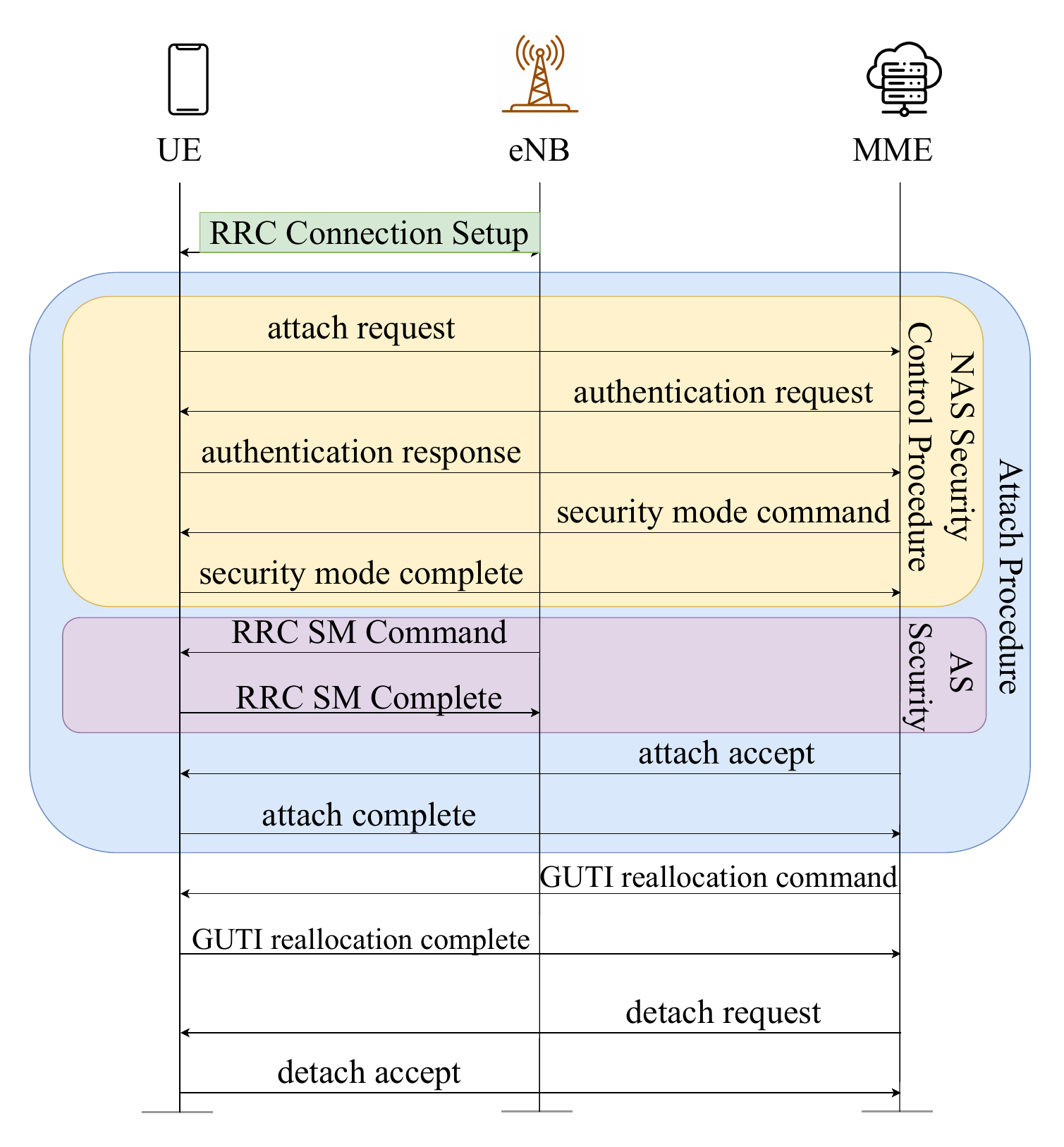}
    \vspace{-0.5cm}
    \caption{Regular flow of NAS and RRC messages in LTE protocol.}
    \label{fig:lte_overview}
    \vspace{-0.65cm}
\end{figure}
\subsection{4G Long Term Evolution (LTE) Protocol}

The 4G LTE is currently the most widely used cellular technology globally. 
The user devices on LTE are called User Equipment (UE), which includes identity information and cryptographic keys. On the network side, eNodeB acts as the base station that provides the radio connection. Moreover, the core network consists of several important components that provide different functionalities in LTE. Among them, the Mobility Management Entity (MME) is responsible for UE's attaching to the network.

LTE protocol stack, at the lowest level, has the Physical Layer. Sequentially, the next layers are the Medium Access Layer (MAC), Radio Link Control (RLC), Radio Resource Control (RRC), Packet Data Convergence Control (PDCP), and Non Access Stratum (NAS) layer, respectively. Among these layers, RRC and NAS are responsible for establishing radio connections, UE's attaching to the network and updating the location of the UE within the network. In this work, we primarily focus on the attach procedure.

The attach procedure starts with UE sending an \attachreq message to the network. Then, after proper authentication via the exchange of \authreq and \authresp messages, the MME establishes the NAS security mode control procedure via the exchange of \smcmd and \smcomplete messages and finally completes the attachment via the exchange of \attachacc and \attachcomp messages. In the RRC layer, AS security context messages are established via the exchange of \rrcsmcmd and \rrcsmcmplt messages between the UE and the base station (eNB). Additionally, the MME can identify information from the UE through \idreq message, to which the UE should reply as per the specification, based on the state and identity type requested. After attachment, the MME also periodically changes the temporary identifier of the UE via the exchange of \gutiral and \gutiralc messages. A sequence diagram of the messages is given in Figure \ref{fig:lte_overview}.

%\ishtiaq{please use correct formats for the messages and revise as necessary.}

\begin{figure}[ht]
    \centering
    \includegraphics[width=0.8\linewidth]{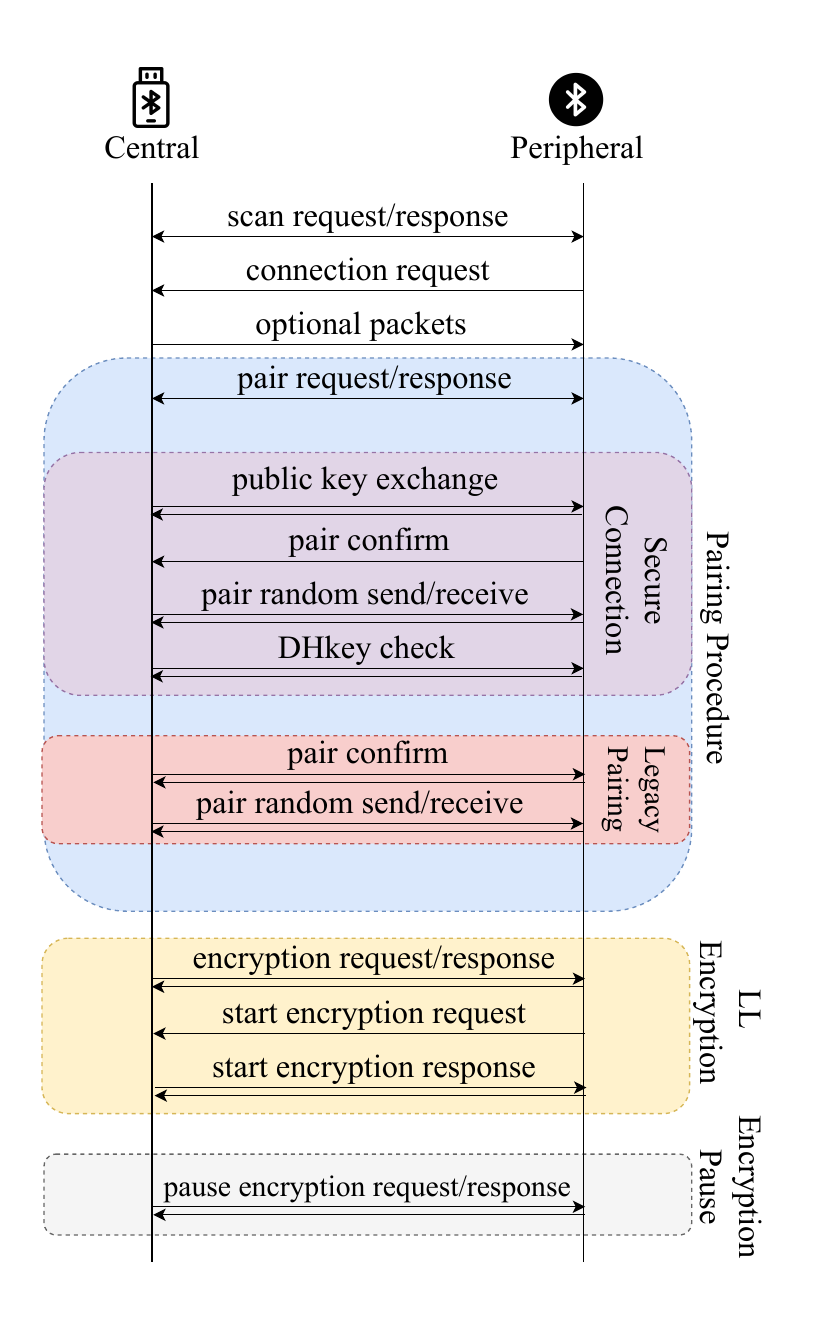}
    \vspace{-0.9cm}
    \caption{Regular flow of BLE protocol.}
    \label{fig:fig-ble-overview}
    \vspace{-0.7cm}
\end{figure}

\subsection{Bluetooth Low Energy (BLE) Protocol}
For short-range communication, Bluetooth is one of the most popular technologies, being widely adopted in numerous devices, including smartphones, IoT devices, and peripherals. 
Among different Bluetooth technologies, BLE provides an energy-efficient communication protocol appropriate for low-cost, energy-restricted devices.

The BLE protocol stack has two distinct subsystems-- \textit{controller} and \textit{host}. The controller subsystem includes the physical layer, the link layer (LL), and the host-controller interface. On the other hand, the host includes Logical Link Control and Adaptation Layer Protocol (L2CAP), Security Manager Protocol (SMP), Attribute Protocol (ATT), Generic Attribute Protocol (GATT), and Generic Access Protocol (GAP).

A BLE pairing and bonding process (As shown in Figure \ref{fig:fig-ble-overview}) starts by establishing the Link Layer connection via the \scanreq, \scanresp and \connectionreq message between the central and peripheral devices and exchanging several optional packets (e.g., \versionreq, \lengthreq, \mtureq message) to determine different connection parameters. After that, the pairing procedure takes place with \pairreq and \pairresp messages. Depending on the type of pairing--- legacy pairing or secure connections--- different sets of messages are exchanged. For legacy pairing, \pairconfirm and \pairrandom messages are exchanged. For secure connections, in addition to these messages, \keyexchange and \dhkeycheck messages are exchanged. 
Finally, the Link Layer encryption is established via the exchange of \encreq and \encresp messages, and the devices' pairing and bonding are completed. The encryption can also be paused between two paired devices via \pauseencreq and \pauseencresp messages. 
%Figure \ref{fig:fig-ble-overview} shows a regular flow of the BLE protocol.

%\ishtiaq{please use correct formats for the messages and revise as necessary.}
\section{Algorithm For \system}
\begin{algorithm}[h]
%\vspace{-0.3cm}
\caption{Approach of \system}
\label{algo:proteus_approach}
\footnotesize
\begin{algorithmic}[1]
\Require 
    \Statex $\mathcal{I}_P$ : Implementation under test
    \Statex $\mathcal{M}$: Guiding PSM
    \Statex $t$: Max number of traces that can be tested within testing budget
    \Statex $\Phi$: Set of desired properties to test 
    \Statex $\mu$: Mutation Budget
    \Statex $\lambda$: Length Budget
\Ensure A list of traces followed by $\mathcal{I}_P$ violating any security property $\phi \in \Phi$
\Procedure{TestProtocol}{}%{I, M, $\Phi$, MaxMut, MaxLen, MaxSkeleton}
    \State $\Psi$ = $\emptyset$ \Comment{Violating Skeletons For All Properties}
    \For{each $\phi \in \Phi$}
        \State $\Psi$ = $\Psi$ $\cup$ GetSkeletons($\phi$) \Comment{\ltlparser}
    \EndFor
    \State $\mathcal{T}_{all}$ = $\emptyset$ \Comment{Set of all traces to test}
    \For{each $\sigma_v \in \Psi$}
        \State $\mathcal{T}_{all}$ = $\mathcal{T}_{all} \cup $ GetTraceSet($\mathcal{M}$, $\sigma_v$, $q_{init}$, $\mu$, $\lambda$) \Comment{\dpalgo}
    \EndFor
    \State numTraceTested = 0
    \While{numTraceTested $\leq$ t} \Comment{\dispatcher}
        \State $\gamma$ = SelectTraceToTest($\mathcal{T}_{all}$) \Comment{Scheduler}
        \State $\mathcal{R}$ = ExecuteTrace($\mathcal{M}$, $\mathcal{I}_{P}$, $\gamma$) \Comment{Execute Trace and Observe Response}
        \If{$\gamma$ violates any security property $\phi$}
            \State Report $\gamma$ violates security property $\phi$
        \EndIf
        \State StoreFeedback($\gamma$, $\mathcal{R}$)
        \State numTraceTested  = numTraceTested + 1
    \EndWhile
\EndProcedure    
\end{algorithmic}
%\vspace{-0.2cm}
\end{algorithm}

\section{\system{}'s Effectiveness and Efficiency with Incorrect PSM}
\label{sec:eval_incorrect_fsm}

We also test what would happen in case of incorrect behavior in the guiding PSM due to human errors during its construction. For this experiment, we intentionally altered our guiding PSM for LTE, and deleted the transition for \smcmd / \smcomplete. We run with the same property discussed in the previous section (\S\ref{sec:eval_performance_proteus}). We find that with mutation budget 1 and length budget 8 there is no trace generated, since now from the guiding PSM at least 2 mutations are required to satisfy $\sigma_v$, one to trigger the replayed \gutiral message and one to place the \smcmd / \smcomplete observation. Consequently, if we run the same example with mutation budget 3 and length budget 9, we get 1920 test traces which also contain a trace that generates the counterexample that is accepted by a practical protocol implementation with the vulnerability. Without the error, only 121 traces are generated with mutation budget 2 and length budget 8 that contains the vulnerability. Thus we conclude that even with the introduction of the error \system can detect the vulnerability, but the error affects its efficiency and we have to run \dpalgo with a higher mutation and length budget to detect it.

\begin{table}[]
    \caption{List of tested LTE devices and identified vulnerabilities. $^*$: The device is known to have the vulnerability.}
    \vspace{-0.3cm}
    \centering
    %\renewcommand{\arraystretch}{1}
    % \fontsize{6}{6}\selectfont
    \resizebox{1\linewidth}{!}{
    \begin{tabular}{|c|c|c|p{4.0cm}|} 
        \hline
        \textbf{Device Name} & \textbf{SoC Model} & \textbf{Baseband Version} & \textbf{Identified Vulnerabilities} \\
	\hline
        Huawei P40 Pro & Kirin 990 5G & 21C93B373S000C000 & L-E9 \\
  	\hline
	Hisense F50+ & Tiger T7510 & 5G\_MODEM\_20C\_W21.12.3\_P5 & L-E3, L-E4  \\
        \hline
	  Galaxy S21 & Exynos 2100 & G991BXXU5CVF3 & L-E1, L-E2, L-E7 \\
	\hline
	  Pixel 6 & Google Tensor & g5123b-116954-230524-B-10194842 & L-E1, L-E2, L-E7 \\
  	\hline 
	Pixel 7 & Google Tensor G2 & g5300q-230626-230818-B-10679446 & L-E2, L-E5, L-E7 \\
        \hline
        Xperia 10 IV & Snapdragon 695 & strait.gen-01223-04 & L-E6, L-E8, L-O1 \\
        \hline
        HTC One E9+ & Helio X10 & 1.1506V24P22T34.2103.0805\_AD5W & L-E1, L-E2 \\
  	\hline
        Nexus 6P & Snapdragon 810 & angler-03.88 & L-E3$^*$, L-E6, L-E8, L-O1 \\
  	\hline
        Galaxy A71 & Snapdragon 730 & A715WVLU4DVI3 & L-E6, L-E8, L-O1 \\
  	\hline
        Pixel 3a & Snapdragon 670 & g670-00042-200421-B-6414611 & L-E3$^*$, L-E6$^*$, L-E8, L-O1 \\
  	\hline 	
        Huawei P8 Lite & Kirin 620 & 22.300.09.00.00 & L-E3, L-E4, L-E7 \\
  	\hline
	%Standard i/o implementation & -- & inspired by ~\cite{lteinspector} formal model &2654 &23290 &1.12 &17&450\\
	%\hline
    \end{tabular}
    }
    \label{tab:device_info_lte}
\end{table}
%\vspace{-0.5cm}
\def\arraystretch{1}
\begin{table}[h]
    \caption{List of tested BLE devices and identified vulnerabilities. $^*$: The device is known to have the vulnerability.}
    \vspace{-0.2cm}
    \centering
    %\renewcommand{\arraystretch}{1}
    % \fontsize{6}{6}\selectfont
    \resizebox{1\linewidth}{!}{
    \begin{tabular}{|c|c|c|p{9cm}|} 
        \hline
        \textbf{Device Name} & \textbf{Vendor} & \textbf{BLE Version} & \textbf{Identified Vulnerabilities}  \\
	\hline
        DT100112 & Microchip & 4.2 & B-E1$^*$, B-E3$^*$, B-E4, B-E5$^*$, B-E7, B-E8$^*$, B-E9, B-O1, B-O3 \\ 
  	\hline
        ESP32-C3 & Espressif & 5.0 & B-E3$^*$, B-E7, B-E10, B-E11, B-I1, B-O3 \\
        \hline
	Hisense F50+ & Hisense & 4.2 & B-E3, B-E4, B-E10, B-E11, B-O3 \\ 
        \hline
	  Galaxy S10 & Samsung & 5.0 & B-E3, B-E4, B-E7, B-I1, B-O1, B-O3 \\
	\hline
        Galaxy S6 & Samsung & 4.1 & B-E1$^*$, B-E3$^*$, B-E4, B-E7, B-E11, B-O1, B-O3 \\
	\hline
        Galaxy A22 & Samsung & 5.0 & B-E2, B-E3, B-E4, B-E6, B-E7, B-E10, B-I1, B-O1, B-O2, B-O3 \\
  	\hline
	  Pixel 6 & Google & 5.2 & B-E4, B-E7, B-E11, B-I1, B-O1, B-O3 \\
  	\hline 
	Pixel 7 & Google & 5.2 & B-E3, B-E4, B-E7, B-E11, B-I1, B-O1, B-O3 \\
        \hline
        Xperia 10 IV & Sony & 5.1 & B-E3, B-E7, B-O1, B-O3 \\
        \hline
 	OPPO Reno7 Pro 5G & OPPO & 5.2 & B-E1, B-E2, B-E3, B-E4, B-E6, B-E7, B-E10, B-E11, B-I1, B-O1, B-O2, B-O3\\
  	\hline
        Motorola Edge+ (2022) & Motorola & 5.2 & B-E3, B-E7, B-I1, B-O1, B-O3 \\
  	\hline
       \multirow{2}{*}{Laptop} & \multirow{2}{*}{Lenovo} & BTstack 1.5.6.3 & \multirow{2}{*}{B-E7, B-E9, B-I1, B-O3} \\
        &  & with BLE 5.2 & \\
  	\hline
        
	%Standard i/o implementation & -- & inspired by ~\cite{lteinspector} formal model &2654 &23290 &1.12 &17&450\\
	%\hline
    \end{tabular}
    }
    \label{tab:device_info_BLE}
\end{table}
%\vspace{-0.8cm}
%\input{tables/securityProperty}
%\vspace{-0.3cm}
%\section{Glossary of Symbols Used}
%\label{sec:glossary}
\vspace{-0.8cm}
\begin{table}[]
\caption{Glossary of symbols used.}
\vspace{-0.2cm}
\resizebox{1\linewidth}{!}{
    \begin{tabular}{|m{1.25cm}|m{9cm}|}
    %\begin{tabular}{|l|l|}
    \hline
    \textbf{Symbol}      & \textbf{Meaning} \\ \hline
    $\mathcal{M}$        & Guiding protocol state machine (PSM) \\ \hline
    $\mathcal{Q}$        & The set of states in a PSM \\ \hline
    $\Sigma$             & The alphabet of input symbols \\ \hline
    $\Lambda$            & The alphabet of output symbols \\ \hline
    $q$                  & A single protocol state. $q_{init}$ denotes the initial protocol state, $q_{c}$ denotes the current protocol state, $q_{n}$ denotes the destination/next protocol state of a transition \\ \hline
    $\mathcal{R}$        & The set of transitions in a PSM \\ \hline
    $\alpha$             & Input symbol of a transition in a PSM \\ \hline
    $\gamma$             & Output symbol of a transition in a PSM \\ \hline
    $\pi$                & A single trace/trace skeleton. $\pi^{a}$ denotes an abstract trace/test skeleton, whereas $\pi^{c}$ denotes an instantiated trace. $\pi_i$ denotes the $i^{th}$ element or observation of a trace. \\ \hline
    $\phi$               & A single security property \\ \hline
    $\Phi$               & A set of security properties \\ \hline
    $\mu$                & Mutation budget (maximum number of perturbations \newline to be applied in a good protocol flow) \\ \hline
    $\lambda$            & Length budget (maximum size of a test case or length of a message sequence) \\ \hline
    $\beta$              & Testing budget. Is a combination of ($\mu,\lambda$) \\ \hline
    $l_j$                & $j^{th}$ observation in the testing skeleton \\ \hline
    \pimp                & Target protocol implementation under test \\ \hline
    \end{tabular}
}
\label{tab:glossary_symbols}
\end{table}

\end{document}